\documentclass[11pt,a4paper]{article}
\pdfoutput=1

\usepackage{jheppub-nosort,amsthm}

\usepackage{amsmath,amssymb,amsthm,amscd}
\usepackage{tikz}
\tikzset{node distance=2cm, auto}
\usepackage{epsfig}
\usepackage{psfrag,comment}
\usepackage{graphicx}
\usepackage{caption}
\usepackage{subcaption}
\usepackage{mathrsfs}
\usepackage{xcolor}
\usepackage[normalem]{ulem}
\usepackage{soul} 

\usepackage{multirow}
\usepackage{booktabs} 
\newcommand{\ra}[1]{\renewcommand{\arraystretch}{#1}}
\usepackage{dcolumn} 
\usepackage{longtable} 
\usepackage{pdflscape} 
\usepackage{pdfpages} 

\newcommand{\CB}{{\cal B}}

\newcommand{\CF}{{\cal F}}

\newcommand{\CH}{{\cal H}}

\newcommand{\CN}{{\cal N}}
\newcommand{\CO}{{\cal O}}

\newcommand{\CQ}{{\cal Q}}

\newcommand{\CS}{{\cal S}}

\def\BN{{\mathbb N}}
\def\BZ{{\mathbb Z}}

\def\BC{{\mathbb C}}
\def\BP{{\mathbb P}}

\newcommand{\be}{\begin{equation}}
\newcommand{\ee}{\end{equation}}
\newcommand{\ba}{\begin{aligned}}
\newcommand{\ea}{\end{aligned}}
\newcommand{\bea}{\begin{eqnarray}}
\newcommand{\eea}{\end{eqnarray}}
\newcommand{\bean}{\begin{eqnarray*}}
\newcommand{\eean}{\end{eqnarray*}}

\newcommand{\p}{\partial}

\def\r{\right\rangle}

\def\1{\mathbf{1}}
\def\0{|\1\r}
\def\im{{\mathbb{I}}{\mathrm{m}}}
\def\re{{\mathbb{R}}{\mathrm{e}}}

\newcommand{\tr}{{\mathrm{Tr}}}
\newcommand{\rme}{{\mathrm{e}}}
\newcommand{\rmi}{{\mathrm{i}}}
\newcommand{\rmd}{{\mathrm{d}}}
\newcommand{\rd}{{\mathrm{d}}}

\def\XXint#1#2#3{{\setbox0=\hbox{$#1{#2#3}{\int}$}
     \vcenter{\hbox{$#2#3$}}\kern-.5\wd0}}

\newcommand{\figref}[1]{Fig.~\protect\ref{#1}}

\newtheorem{thm}{Theorem}

\newtheorem{cor}{Corollary}

\def\IP{\mathbb{P}}
\def\IR{\mathbb{R}}
\def\IZ{\mathbb{Z}}
\newcommand{\mJ}{{\mathsf{J}}}
\newcommand{\mx}{{\mathsf{x}}}
\newcommand{\my}{{\mathsf{y}}}

\newcommand{\mb}{{\mathsf{b}}}
\newcommand{\map}{{\mathsf{p}}}
\newcommand{\mq}{{\mathsf{q}}}

\newcommand{\mO}{{\mathsf{O}}}
\newcommand{\ri}{{\rm i}}
\newcommand{\rem}{{\rm e}}
\newcommand{\fad}{\operatorname{\Phi}_{\mathsf{b}}}
\newcommand{\mypsi}[2]{\operatorname{\Psi}_{#1,#2}}

\definecolor{pert}{cmyk}{0.5, 0.39, 0., 0.3}
\definecolor{1inst}{cmyk}{0.5, 0., 1., 0.4}
\definecolor{2inst}{cmyk}{0., 0.255, 0.75, 0.1}
\definecolor{3inst}{cmyk}{0., 0.6643, 0.73, 0.27}
\newcommand{\bthm}{\begin{thm}}
\newcommand{\ethm}{\end{thm}}
\newcommand{\bprop}{\begin{prop}}
\newcommand{\eprop}{\end{prop}}
\newcommand{\bcor}{\begin{cor}}
\newcommand{\ecor}{\end{cor}}
\newcommand{\bdefn}{\begin{defn}}
\newcommand{\edefn}{\end{defn}}
\newcommand{\brem}{\begin{rem}}
\newcommand{\erem}{\end{rem}}

\newcommand{\G}{\Gamma}

\def\1{\mathbf{1}}

\def\XXint#1#2#3{{\setbox0=\hbox{$#1{#2#3}{\int}$}
     \vcenter{\hbox{$#2#3$}}\kern-.5\wd0}}

\newcommand{\Pol}{{\text{Pol}}}

\makeatletter
\let\oldabs\abs
\def\abs{\@ifstar{\oldabs}{\oldabs*}}

\newcommand{\txK}{\ensuremath{\text{K}}}
\newcommand{\txc}{\ensuremath{\text{c}}}

\newcommand{\txeff}{\ensuremath{\text{eff}}}
\newcommand{\txRC}{\ensuremath{\text{rc}}}

\newcommand{\instsect}[2]{(#1)\langle #2 \rangle}

\newcommand{\FPtwo}{F_{\BP^2}}
\newcommand{\hFPtwo}{\CF_{\BP^2}}

\newcommand{\Fn}[1]{F^{(#1)}}
\newcommand{\Fng}[2]{F^{(#1)}_{#2}}
\newcommand{\FnAg}[3]{F^{\instsect{#1}{#2}}_{#3}}
\newcommand{\fng}[2]{f^{(#1)}_{#2}}

\newcommand{\hFn}[1]{\CF^{(#1)}}

\newcommand{\hFnA}[2]{\CF^{\instsect{#1}{#2}}}
\newcommand{\hFnAg}[3]{\CF^{\instsect{#1}{#2}}_{#3}}
\newcommand{\hFngA}[3]{\CF^{(#1)}_{#2;#3}}
\newcommand{\hFnAgA}[4]{\CF^{\instsect{#1}{#2}}_{#3;#4}}
\newcommand{\hFnfA}[2]{\CF^{(#1)}_{;#2}}
\newcommand{\hFnAfA}[3]{\CF^{\instsect{#1}{#2}}_{;#3}}

\newcommand{\thFnA}[2]{\widetilde{\CF}^{\instsect{#1}{#2}}}
\newcommand{\thFnAg}[3]{\widetilde{\CF}^{\instsect{#1}{#2}}_{#3}}

\newcommand{\Szz}{S^{zz}}
\newcommand{\hSzzA}[1]{\CS^{zz}_{;#1}}

\newcommand{\stk}[1]{S_{#1,1}}

\newcommand{\stpwn}[1]{b^{(#1)}}
\newcommand{\stpwnA}[2]{b^{\instsect{#1}{#2}}}

\newcommand{\sgm}[1]{\sigma_{#1}}
\newcommand{\tsgm}[1]{\widetilde{\sigma}_{#1}}

\makeatletter
\newsavebox\myboxA
\newsavebox\myboxB
\newlength\mylenA

\newcommand*\widebar[2][0.75]{%
    \sbox{\myboxA}{$\m@th#2$}%
    \setbox\myboxB\null
    \ht\myboxB=\ht\myboxA%
    \dp\myboxB=\dp\myboxA%
    \wd\myboxB=#1\wd\myboxA
    \sbox\myboxB{$\m@th\overline{\copy\myboxB}$}
    \setlength\mylenA{\the\wd\myboxA}
    \addtolength\mylenA{-\the\wd\myboxB}%
    \ifdim\wd\myboxB<\wd\myboxA%
       \rlap{\hskip 0.8\mylenA\usebox\myboxB}{\usebox\myboxA}%
    \else
        \hskip -0.5\mylenA\rlap{\usebox\myboxA}{\hskip 0.5\mylenA\usebox\myboxB}%
    \fi}
\makeatother

\soulregister{\emph}{1}

\newcommand{\dblunderline}[2]{\uwave{\underline{#1}#2}}

\title{Resurgence Matches Quantization}

\author[a]{Ricardo~Couso-Santamar\'\i a,}
\affiliation[a]{CAMGSD, Departamento de Matem\'atica, Instituto Superior T\'ecnico,\\ Universidade de Lisboa, Av. Rovisco Pais 1, 1049-001 Lisboa, Portugal\\}
\emailAdd{santamaria@math.tecnico.ulisboa.pt}

\author[b]{Marcos~Mari\~no,}
\affiliation[b]{D\'epartement de Physique Th\'eorique \& Section de Math\'ematiques,\\ Universit\'e de Gen\`eve, Gen\`eve, CH-1211 Switzerland\\}
\emailAdd{marcos.marino@unige.ch}

\author[b,a]{Ricardo~Schiappa\,}
\emailAdd{schiappa@math.tecnico.ulisboa.pt}

\abstract{
The quest to find a nonperturbative formulation of topological string theory has recently seen two unrelated developments. On the one hand, via quantization of the mirror curve associated to a toric Calabi--Yau background, it has been possible to give a nonperturbative definition of the topological-string partition function. On the other hand, using techniques of resurgence and transseries, it has been possible to extend the string (asymptotic) perturbative expansion into a transseries involving nonperturbative instanton sectors. Within the specific example of the local $\BP^2$ toric Calabi--Yau threefold, the present work shows how the Borel--Pad\'e--\'Ecalle resummation of this resurgent transseries, alongside occurrence of Stokes phenomenon, matches the string-theoretic partition function obtained via quantization of the mirror curve. This match is highly non-trivial, given the unrelated nature of both nonperturbative frameworks, signaling at the existence of a consistent underlying structure.
}


\arxivnumber{1610.06782}

\begin{document}

\maketitle

\vfill

\eject

\allowdisplaybreaks

\section{Introduction}\label{sec:intro}

Historically, string theory is defined by a perturbative expansion. This expansion turns out to be asymptotic, \textit{e.g.}, \cite{gp88}, with zero radius of convergence; signaling at the existence of a variety of nonperturbative effects beyond the original perturbative definition, \textit{e.g.}, \cite{s90}. Nonetheless, in spite of decades of research and enormous progress, a fully general nonperturbative definition of string theory is still lacking. One simplified setting where this question may be addressed, and where significant progress has been achieved recently, is that of topological string theory. This is the scenario we shall address in this paper, in the special case of toric Calabi--Yau backgrounds.

Asymptotic expansions alone cannot define the topological-string free energy. Their lack of convergence asks for resummation methods\footnote{In principle, Borel resummation; in practice its numerical implementation as Borel--Pad\'e resummation.} but even these are not enough due to Stokes phenomenon: different asymptotic expansions hold, in different regions of the parameters, leading to distinct resummations. All this richness is single-captured by the extension of the original asymptotic expansion into a (resurgent) transseries, which includes non-analytic data (see, \textit{e.g.}, \cite{abs16} for an introduction and a complete list of references). This resulting transseries finally yields a nonperturbative framework to reconstruct the original function we wish to uncover (albeit iteratively, in the case of non-linear problems). For the case of the topological-string free energy, the general construction of its resurgent transseries---out of a nonperturbative extension of the holomorphic anomaly equations---was set-up in \cite{cesv13}; with the explicit example of the local $\BP^2$ toric Calabi--Yau threefold being fully worked out in \cite{cesv14}. These results were obtained based upon earlier stringy constructions \cite{m06, msw07, m08, msw08, ps09, gikm10, kmr10, dmp11, asv11, sv13}, and have since led to a few further developments, \textit{e.g.}, \cite{c15, csv16}. In principle, this construction allows us to obtain fully nonperturbative results for the string-theoretic free energy, at any value of the string coupling constant.

A different but very much related question is to ask if, instead of having a \textit{method} to obtain nonperturbative results out of somewhat perturbative data, 
one may have a \textit{definition} of the theory which is nonperturbative from scratch. This is the spirit of the familiar large $N$ dualities \cite{m97}, where the string genus expansion is regarded as a (perhaps complicated) asymptotic expansion of some simpler theory---which itself then becomes the nonperturbative definition of the problem. As a framework for quantum gravity, it is somewhat expectable that string theory should be associated to some sort of quantization of the background, spacetime geometry. As recently uncovered, such an expectation turns out to be realized when addressing topological strings on toric Calabi--Yau backgrounds---where the ``spacetime'' geometry is essentially encoded in the associated mirror manifold, itself built upon an algebraic curve. In this case, the quantization of this ``mirror curve'' leads to a simple one-dimensional, self-adjoint quantum mechanical operator, defined on the real line. Based upon earlier constructions \cite{adkmv03, ns09, mm09, acdkv11, mp11, hmo12b, cm12, hmo13, hmmo13, km13, hw14, cgm14}, it was proposed in \cite{ghm14a} that there is a precise correspondence between the spectral theory of operators obtained by quantizing the mirror curve, and topological string theory on the toric Calabi--Yau geometry. This proposal has since led to many recent developments, \textit{e.g.}, \cite{ghm14b, km15, mz15, kmz15, wzh15, gkmr15, cgm15, h15b, hm15, fhm15, oz15, bgt16, hkt16, k16, g16, s16, mz16, swh16, cgm16, gg16} (see \cite{m15} for an introduction). In addition, the topological string/spectral theory correspondence of \cite{ghm14a} provides a nonperturbative definition of the topological-string partition function on toric Calabi--Yau geometries. According to 
this correspondence, the ``fermionic'' spectral traces of these quantum-mechanical operators, which are well-defined functions, have a natural 't~Hooft-like limit which results in an asymptotic expansion yielding the weakly-coupled topological-string genus expansion. 
 
It is well-known that, in many quantum systems, nonperturbative quantities are exactly given by resummations of transseries. Examples include the energy levels of the quartic anharmonic oscillator \cite{ggs70} and of the double-well potential \cite{zj05}, as well as the partition functions of some non-critical string theories \cite{m08} and matrix models \cite{csv15}. In these examples, the existence of a transseries representation means that the nonperturbative object can be ``decoded'' in terms of semi-classical information. It is then natural to ask whether the nonperturbative topological-string free energy, as defined in terms of spectral theory in \cite{ghm14a}, can be expressed as the resummation of the general transseries constructed in \cite{cesv13, cesv14}. In this paper we shall obtain a \textit{precise match} between these two quantities, once Stokes phenomenon is taken into account. Due to the independent nature of these two constructions, this match is in fact highly non-trivial, and it implies a consistent underlying structure common to both approaches. As a consequence, our results will hopefully also stand as a benchmark for future tests of any other (independent) approach to the nonperturbative definition of the topological string, which will need to find their place within such underlying setting. We begin in section~\ref{sec:exact} with a description of the quantization approach, and proceed in section~\ref{sec:transseries} with a description of the transseries construction. The match is then established via resummation in section~\ref{sec:resummation}. In particular, this match validates (although it does not prove) the semi-classical data in \cite{cesv13, cesv14, c15} as the complete set of nonperturbative (instanton) semi-classical data in the underlying theory. We close with a short discussion in section~\ref{sec:discussion}.

\section{A Nonperturbative Definition of Topological Strings}\label{sec:exact}

A nonperturbative definition for the topological-string free energy, in the conifold frame and in the case of toric Calabi--Yau (CY) threefolds, was proposed in \cite{ghm14a, mz15}. This definition is based upon the quantization of the mirror curve to the CY, as originally suggested in \cite{adkmv03}, and is explicitly calculable. We shall review this definition in the following, focusing for simplicity on toric del~Pezzo CY threefolds, whose mirror curve has genus one (the general construction, in the case 
of mirror curves of higher genus, can be found in \cite{cgm15}).

\subsection{Topological Strings on Toric Calabi--Yau Threefolds}

Recall that toric del~Pezzo CYs are defined as the total space of the canonical bundle on a toric (almost) del~Pezzo surface $S$,
\be
\label{dP}
X=\CO(K_S) \rightarrow S,
\ee
\noindent
and they are sometimes called ``local $S$''. The simplest example of a local del~Pezzo occurs when $S=\IP^2$. This is the local $\IP^2$ geometry that we shall focus on in most of this paper. By standard results in toric geometry (see for example \cite{hkp13}), toric almost del~Pezzo surfaces can be classified by reflexive polyhedra in two dimensions. The polyhedron $\Delta_S$ associated to a surface $S$ is the convex hull of a set of two-dimensional vectors 
\be
\label{vector}
\nu^{(i)}=\left(\nu^{(i)}_1, \nu^{(i)}_2\right), \qquad i=1, \cdots, s+2, 
\ee
\noindent
together with the origin. In this case, the associated CY in \eqref{dP} has $s$ K\"ahler parameters. As shown in \cite{kkv96, ckyz99}, the mirror geometry to \eqref{dP} can be encoded in an algebraic curve written in exponentiated variables, $\rem^x$ and $\rem^y$, which is known as the \textit{mirror curve}. In the case of local del~Pezzo CYs, the curve has genus one. The $s$ complex parameters characterizing this curve can thus be divided into two types: one ``true'' modulus, $\kappa$, and a set of ``mass'' parameters, $\xi_i$, $i=1, \ldots, s-1$ \cite{hkp13, hkrs14}. The equation for the mirror curve can then be written in a canonical form,
\be
\label{ex-W}
W (\rem^x, \rem^y) = \CO_S(x,y) + \kappa = 0,
\ee
\noindent
where the function $\CO_S(x,y)$ is given by 
\be
\label{coxp}
\CO_S (x,y)=\sum_{i=1}^{s+2} \exp\left( \nu^{(i)}_1 x+  \nu^{(i)}_2 y + f_i(\boldsymbol{\xi}) \right), 
\ee
\noindent
and where $f_i(\boldsymbol{\xi})$ are suitable functions of the parameters $\xi_j$. 

\begin{figure}[t!]
\begin{center}
\includegraphics[scale=0.5]{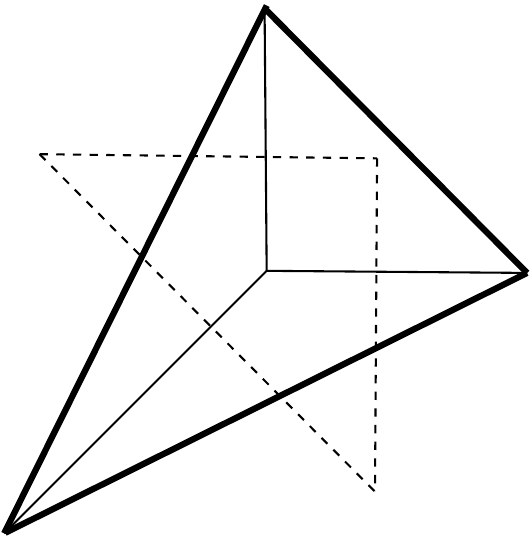}
\end{center}
\caption{The vectors \eqref{p2-vec} defining the local $\IP^2$ geometry, together with the polyhedron $\Delta_{\IP^2}$ (in thick lines) and the dual polyhedron (in dashed lines).}
\label{hut}
\end{figure}

As mentioned, our main example will be local $\IP^2$. For this CY, $s=1$ and there are no mass parameters. The vectors \eqref{vector} are given by 
\be
\label{p2-vec}
\nu^{(1)}=(1,0), \qquad \nu^{(2)}=(0,1), \qquad \nu^{(3)}=(-1,-1),
\ee
\noindent
and we plot these vectors, together with the origin and their convex hull $\Delta_{\IP^2}$, in \figref{hut}. In this case, the function \eqref{coxp} will be given by 
\be
\label{localP2exp}
\CO_{\IP^2}(x,y)= \rem^x+ \rem^y+ \rem^{-x-y}. 
\ee

Given a CY threefold $X$, we shall denote the closed topological-string free energy at genus $g$, and in the large-radius frame, by $F_g(\boldsymbol{t})$, where $\boldsymbol{t}$ is the vector of K\"ahler parameters. These functions are generating functionals for the Gromov--Witten (GW) invariants of $X$, and have the structure:
\be
\label{gzp}
\ba
F_0(\boldsymbol{t}) &= \frac{1}{6} a_{ijk}\, t_i t_j t_k  + \sum_{\boldsymbol{d}}  N_0^{\boldsymbol{d}}\, \rem^{-\boldsymbol{d} \cdot \boldsymbol{t}}, \\
F_1(\boldsymbol{t}) &= b_i\, t_i + \sum_{\boldsymbol{d}} N_1^{\boldsymbol{d}}\, \rem^{-\boldsymbol{d} \cdot \boldsymbol{t}}, \\
F_g(\boldsymbol{t}) &= C_g + \sum_{\boldsymbol{d}} N_g^{\boldsymbol{d}}\, \rem^{-\boldsymbol{d} \cdot \boldsymbol{t}}, \qquad g\ge 2. 
\ea
\ee
\noindent
In these formulae, repeated indices are summed over, and $N_g^{\boldsymbol{d}}$ are the GW invariants of $X$ at genus $g$ and multi-degree $\boldsymbol{d}$. The coefficients $a_{ijk}$ and $b_i$ are cubic and linear couplings characterizing the perturbative genus-zero and genus-one free energies, while $C_g$ is the so-called constant map contribution \cite{bcov93}. The total free energy of the topological string is the formal\footnote{As we shall discuss in the upcoming section~\ref{sec:transseries}, this is an asymptotic series.} series, 
\be
\label{tfe}
F \left(\boldsymbol{t}, g_{\text{s}}\right) \simeq \sum_{g=0}^{+\infty} g_{\text{s}}^{2g-2} F_g(\boldsymbol{t}), 
\ee
\noindent
where $g_{\text{s}}$ is the topological-string coupling constant.

\subsection{The Quantum Mechanical Proposal}

In the case of local del~Pezzo CYs, the ``quantization'' of the mirror curve \eqref{ex-W} is based on the promotion of the function $\CO_S(x,y)$ to an operator, which will be denoted by $\mathsf{O}_S$. To do this, we first promote $x$ and $y$ to self-adjoint Heisenberg operators on the real line, $\mathsf{x}$ and $\mathsf{y}$, satisfying the standard commutation relation
\be
[\mx, \my]= \ri \hbar, \qquad \hbar \in \IR. 
\ee
\noindent
Next, we apply Weyl's quantization to $\CO_S(x,y)$. In particular, this fixes possible ordering ambiguities and leads to an unbounded, self-adjoint operator $\mO_S$ on the Hilbert space $\CH= L^2(\IR)$. It was further conjectured in \cite{ghm14a} that the inverse operator,
\be
\label{rhos}
\rho_S=\mathsf{O}^{-1}_S,
\ee
\noindent
acting on $L^2(\IR)$, is of trace class and positive definite (provided the mass parameters $\xi_i$ satisfy some positivity conditions). This was later proved in \cite{km15, lst15}. With $\rho_S$ of trace class, one can then consider its Fredholm determinant (see for example \cite{s77})
\be
\label{gsd}
\Xi_S(\kappa, \hbar) = \text{det} \left( 1 + \kappa\, \rho_S \right),
\ee
\noindent
and it is a standard result that $\Xi_S(\kappa,\hbar)$ is an entire function of $\kappa$. Its Taylor series around the origin,
\be
\Xi_S(\kappa, \hbar)= 1 + \sum_{N\ge 1} Z_S (N, \hbar)\, \kappa^N, 
\ee
\noindent
defines the \textit{fermionic spectral traces} $Z_S(N, \hbar)$. They can also be obtained from the ``bosonic'' spectral traces, $\tr \rho_S^\ell$, $\ell=1,2,\ldots$, through Fredholm--Plemelj's formula
\be
\label{f-p}
\Xi_S (\kappa, \hbar) = \exp \left\{ -\sum_{\ell=1}^{+\infty} \frac{(-\kappa)^\ell}{\ell}\, \tr\, \rho_S^\ell \right\}.  
\ee

In \cite{ghm14a}, conjectural, exact formulae for both $\Xi_S(\kappa, \hbar)$ and $Z_S(N, \hbar)$ were proposed, in terms of BPS invariants of the local del~Pezzo CY. One of the consequences of these conjectures, emphasized in \cite{mz15}, is that $Z_S(N, \hbar)$ provides a nonperturbative definition of the topological-string partition function. More precisely, the logarithm of the fermionic spectral traces, 
\be
F_S(N, \hbar) = \log Z_S(N, \hbar),
\ee
\noindent
which are manifestly well-defined quantities for all $\hbar \in \IR_{>0}$ and all $N \in \IZ_{>0}$, has an asymptotic expansion which is precisely the genus expansion of the topological string, in the so-called \textit{conifold frame}. This asymptotic expansion is performed in the \textit{'t~Hooft regime}, characterized by 
\be
\label{thooft-lam}
\hbar \rightarrow +\infty, \qquad N \rightarrow +\infty, \qquad \frac{N}{\hbar} \equiv \lambda \quad \text{fixed},
\ee
\noindent
and it has the form
\be
\label{logz-exp}
F_S(N, \hbar) \simeq \sum_{g=0}^{+\infty} \CF_g (\lambda)\, \hbar^{2-2g}.
\ee
\noindent
The functions $\CF_g(\lambda)$ are the topological-string free energies in the conifold frame, while $\lambda$ is a flat coordinate which vanishes at the conifold point. The functions $\CF_g(\lambda)$ can be obtained from the large-radius free energies by performing a formal Laplace transform, as explained in \cite{abk06}. We shall next briefly review this procedure, taking into account the precise normalizations and conventions required by the conjecture of \cite{ghm14a, mz15}. Additional details can be found in, \textit{e.g.}, \cite{m15}. We will also focus on the situation relevant for local $\IP^2$, so we shall assume that we have a trivial dependence on the mass parameters, and therefore a single K\"ahler parameter $t$. 

In the conjecture of \cite{ghm14a, mz15}, one has to introduce a non-trivial B-field, $B$, in the standard definition of the topological-string free energies. This B-field is determined by the geometry of the CY \cite{hmmo13, ghm14a, swh16}. We then define
\be
\widehat{F}_g ( t) = F_g ( t+ \ri\pi B). 
\ee
\noindent
Another required ingredient in the theory, developed in \cite{ghm14a, cgm15}, is a function $A(\hbar)$ which has an asymptotic expansion 
\be
\label{ah}
A ( \hbar) \simeq \sum_{g=0}^{+\infty} A_g\, \hbar^{2-2g}. 
\ee
\noindent
This function has been conjectured for various geometries, in particular for local in $\IP^2$ \cite{ghm14a}. It is also convenient to introduce the following functions, which are closely related to the topological-string free energies,  
\be 
\label{gen-J-as}
\ba
\mJ_0 (t) &= \frac{1}{16\pi^4} \left( \widehat{F}_0 (t) + 4\pi^2\, b^{\text{NS}} t \right) +  A_0, \\
\mJ_1 (t) &= A_1 + \widehat{F}_1 (t), \\
\mJ_g (t) &= A_g + \left( 4\pi^2 \right)^{2g-2} \left( \widehat{F}_g (t)- C_g \right), \qquad g\ge 2. 
\ea
\ee
\noindent
The coefficient $b^{\text{NS}}$ should not be confused with the one appearing in \eqref{gzp}. Instead, it can be determined from the perturbative part of the so-called Nekrasov--Shatashvili (NS) free energy; see \cite{ghm14a, cgm15} for more details and examples. Let us further introduce the formal power series
\be
\label{jts}
\mJ^{\text{TS}} ( t, \hbar ) \simeq \sum_{g=0}^{+\infty} \mJ_g (t)\, \hbar^{2-2g}. 
\ee
\noindent
This is, up to the small modifications in \eqref{gen-J-as}, the total topological-string free energy in the large-radius frame. It also follows from the above 
formulae that the string coupling $g_{\text{s}}$ is related to $\hbar$ by 
\be
g_{\text{s}} = \frac{4\pi^2}{\hbar}.  
\ee
\noindent
As advertised above, one may then write the topological-string free energies in the conifold frame, $\CF_g(\lambda)$, via the following formal Laplace transform \cite{abk06}, 
\be
\label{ZAiry}
\exp \left\{ \sum_{g=0}^{+\infty} \CF_g (\lambda)\, \hbar^{2-2g} \right\} \simeq \frac{1}{\ri g_{\text{s}} C}  \int \rmd t\, \exp \left\{ \mJ^{\text{TS}} ( t, \hbar ) - \left( 2 \pi C \right)^{-1} \hbar^2\, \lambda t \right\}. 
\ee
\noindent
Here, the constant $C$ is determined by the mirror map in the large-volume limit. Namely, if we write the mirror curve as in \eqref{ex-W}, and set $\kappa= \rme^\mu$, then the mirror map has the form 
\be
\label{zmu}
t \approx C\mu, \qquad \mu \gg 1. 
\ee

The conifold free energies $\CF_g(\lambda)$ may now be computed by doing a formal, saddle-point approximation to the above Laplace transform \eqref{ZAiry} for large $\hbar$. This saddle point is given by 
\be
\label{saddle}
\lambda = \frac{C}{8\pi^3} \left( \frac{\partial\widehat{F}_0}{\partial t} + 4\pi^2\, b^{\text{NS}} \right).
\ee
\noindent
It follows from this equation that the 't~Hooft parameter is proportional to a period, \textit{i.e.}, it is a flat coordinate on moduli space, as claimed above. The conifold point is given by 
\be
\lambda=0. 
\ee
\noindent
The genus-zero free energy in the conifold frame, $\CF_0(\lambda)$, is given by a Legendre transform,
\be
\CF_0 (\lambda) = \mJ_0 (t) - (2 \pi C)^{-1} \lambda t. 
\ee
\noindent
In practice, the higher-order corrections $\CF_g(\lambda)$ are computed by solving the holomorphic anomaly equations in the conifold frame, as in \cite{hkr08}. 

It should be noted that the formula \eqref{ZAiry} is the asymptotic expansion of an \textit{exact}, conjectural formula for the fermionic spectral traces $Z_S(N, \hbar)$. This exact formula makes it possible to compute $Z_S(N, \hbar)$ by using the BPS invariants of the CY. In addition, it extends $Z_S(N, \hbar)$ (which was originally defined only for positive, integer values of $N$), to an \textit{entire} function on the complex plane of the $N$ variable. This will be also useful in the present paper\footnote{Albeit we shall only use such extension to non-integer, yet \textit{real}, values of $N$.}. Similarly, we note that the formal power-series \eqref{jts} is the asymptotic expansion of a well-defined function, the so-called modified grand potential introduced in \cite{hmmo13} for arbitrary toric CY threefolds.

\subsection{Exact Results for Local $\BP^2$}

Our main example in this paper is that of local $\IP^2$, and we shall now specialize the above discussion to this local del~Pezzo CY, following \cite{ghm14a, mz15}. In this case, as we have already remarked, $s=1$ and there are no mass parameters. The several coefficients appearing in \eqref{gzp}, \eqref{gen-J-as} and \eqref{zmu} take the values 
\be
a=\frac{1}{3}, \qquad b=\frac{1}{12}, \qquad b^{\text{NS}}=-\frac{1}{24}, \qquad C=3. 
\ee
\noindent
The constant-map contribution is given by \cite{hkr08}
\be
\label{Cg}
C_g = 3\, (-1)^{g-1}\, c_g, \qquad g\ge 2, 
\ee
\noindent
where
\be
\label{cg}
c_g = \frac{B_{2g}\, B_{2g-2}}{\left(4g\right) \left(2g-2\right) \left(2g-2\right)!}.
\ee
\noindent
In addition, the B-field takes the value $B=1$, and the coefficients $A_g$ appearing in \eqref{gen-J-as} are given by \cite{mz15}
\be
\label{ags}
\ba
A_0 &= \frac{3\zeta(3)}{16\pi^4}, \\
A_1 &= - \frac{1}{12} \log \hbar + \zeta'(-1) + \frac{1}{6} \log 2\pi + \frac{1}{24} \log 3, \\
A_g &= \left(4\pi^2\right)^{2g-2} \left( 3-3^{2-2g} \right) \left(-1\right)^{g-1} c_g, \qquad g\ge 2. 
\ea
\ee

For the vanishing flat-coordinate at the conifold point we shall use the conventions in \cite{cesv14}, namely 
\be
\label{fcc}
t_{\text{c}} (\psi) = \frac{2\pi}{\sqrt{3}} \left( \frac{3\psi}{\G\left(\frac{2}{3}\right)^3}\,\, {}_3 F_2 \left. \left( \frac{1}{3},\frac{1}{3},\frac{1}{3}; \frac{2}{3},\frac{4}{3}\, \right| \psi^3 \right) - \frac{\frac{9}{2}\psi^2}{\G\left(\frac{1}{3}\right)^3}\,\, {}_3 F_2 \left. \left( \frac{2}{3},\frac{2}{3},\frac{2}{3}; \frac{4}{3},\frac{5}{3}\, \right| \psi^3 \right) - 1 \right), 
\ee
\noindent
where $\psi$ is related to the parameter $\kappa$ appearing in the mirror curve \eqref{ex-W} by
\be
\psi = - \frac{\kappa}{3}. 
\ee
\noindent
Another standard parametrization of the CY complex-structure moduli space is 
\be
\label{z-def}
z = \frac{1}{\kappa^3}, 
\ee
\noindent
with $\psi = (-27z)^{-1/3}$. By comparing with \eqref{saddle}, one finds that $\lambda$ is given by\footnote{When comparing to formulae in \cite{mz15}, one has to take into account that the conifold flat-coordinate used in there differs from \eqref{fcc} by an overall multiplicative constant. In addition, many of the formulae in \cite{mz15} have the opposite sign for the complex modulus $z$. This is due to the way the non-trivial B-field was implemented in \cite{mz15}, but for our present purposes we can use the standard parametrization \eqref{fcc}.}, 
\be
\label{'tHooft-tc}
\lambda = \frac{t_{\text{c}} (z)}{4\sqrt{3}\pi^2}. 
\ee

Using these results, one can obtain the genus-$g$ free energies in the conifold frame. Their expansion around $\lambda=0$, worked out in \cite{mz15}, is very useful in order to fix our conventions. One finds
\be
\label{small-lam}
\ba
\CF_0 (\lambda) &= -\frac{c}{6\pi} \lambda + \left( \frac{1}{2} \log\lambda - \frac{3}{4} + \log \frac{2\pi}{3\sqrt[4]{3}} \right) \lambda^2 + \CO(\lambda^3), \\
\CF_1 (\lambda) &= -\frac{1}{12} \log \lambda \hbar + \zeta'(-1) + \CO(\lambda), \\
\CF_g (\lambda) &= \frac{B_{2g}}{2g \left(2g-2\right)} \lambda^{2-2g} + \CO(\lambda), \qquad g\ge 2. 
\ea
\ee
\noindent
The coefficient $c$ in the first line is given by 
\be
c = \frac{9V}{2\pi}, \qquad V = 2\,\im \left( {\rm Li}_2 \left( \rem^{\ri\pi/3} \right) \right). 
\ee
\noindent
Note that this expansion displays a strong version of the gap condition of \cite{hk06}. 

Let us next address the calculation of the fermionic spectral traces $Z(N, \hbar)$ (in the following we suppress the subscript $S=\IP^2$). The main result that allows the calculation of these traces is an explicit expression for the kernel of the trace-class operator $\rho$, obtained in \cite{km15}. In fact, with the same effort, one can consider slightly more general operators of the form 
\begin{equation}
\label{tto}
\mathsf{O}_{m,n} = \rem^{\mathsf{x}} + \rem^{\mathsf{y}} + \rem^{-m\mathsf{x}-n\mathsf{y}}.
\end{equation}
\noindent
Here, $m, n$ are positive, real numbers; and the operator for local $\IP^2$ corresponds to $m=n=1$. These operators \eqref{tto} were called three-term operators in \cite{km15}. In order to write down the explicit kernel of their inverses, let us first introduce some notation. As in \cite{km15}, we shall denote by $\fad(x)$ Faddeev's quantum dilogarithm \cite{f95, fk93}. We define as well 
\be
\label{mypsi-def}
\mypsi{a}{c}(x) = \frac{\rem^{2\pi ax}}{\fad\left(x-\ri (a+c)\right)}. 
\ee
\noindent
Next, introduce normalized Heisenberg operators, $\mq$ and $\map$, satisfying the normalized commutation relation
\be
[\map, \mq] = \left(2\pi\ri\right)^{-1},
\ee
\noindent
and such that they are related to $\mx$ and $\my$ by the linear canonical transformation
\begin{equation}
\mathsf{x} \equiv 2\pi\mathsf{b}\, \frac{(n+1)\mathsf{p}+n\mathsf{q}}{m+n+1},\qquad \mathsf{y} \equiv -2\pi\mathsf{b}\, \frac{m\mathsf{p}+(m+1)\mathsf{q}}{m+n+1}, 
\end{equation}
\noindent
so that $\hbar$ is related to the parameter $\mb$ in the quantum dilogarithm by
\be
\label{b-hbar}
\hbar=\frac{2\pi\mathsf{b}^2}{m+n+1}. 
\ee
\noindent
In the momentum representation associated to $\map$, our usual operator 
\be
\rho_{m,n} = \mO_{m,n}^{-1}
\ee
\noindent
has the integral kernel
\begin{equation}
\label{ex-k}
\rho_{m,n}(p,p') = \frac{\overline{\mypsi{a}{c}(p)}\, \mypsi{a}{c}(p')}{2\mathsf{b} \cosh\left(\pi\frac{p-p'+\ri (a+c-nc)}{\mathsf{b}}\right)}.
\end{equation}
\noindent
In this equation, $a$ and $c$ are given by 
\be
a = \frac{m\mb}{2(m+n+1)}, \qquad c = \frac{\mb}{2(m+n+1)}. 
\ee
\noindent
Using this result, one may calculate the ``bosonic'' spectral traces as multiple integrals,
\be
\tr\, \rho_{m,n}^\ell = \int \rd p_1 \cdots \rd p_\ell\, \rho_{m,n}(p_1, p_2) \cdots  \rho_{m,n}(p_{\ell-1}, p_\ell). 
\ee
\noindent
Now, when $\mb^2$ is of the form $M/N$, with $M$ and $N$ coprime integers, Faddeev's quantum dilogarithm simplifies into an elementary function \cite{gk14}---and it is in principle possible to calculate these integrals exactly, by residues. Using this technique, many results for the spectral traces have been obtained in \cite{km15}. In order to review these results, let us focus again on local $\IP^2$. In this case, the relation between $\hbar$ and $\mb$ is 
\be
\hbar = \frac{2\pi}{3} \mathsf{b}^2.
\ee
\noindent
For $N=1$ one can obtain \cite{km15} the following closed formula for $Z(1,\hbar)$ at arbitrary $\hbar$:
\be
Z(1,\hbar) = \frac{1}{\mathsf{b}} \left|\fad\left( c_{\mathsf{b}}- \frac{\ri\mathsf{b}}{3} \right)\right|^3.
\ee
\noindent
Using the properties of Faddeev's quantum dilogarithm for rational $\mb^2$ established in \cite{gk14}, one finds for example:
\be
\ba
Z\left(1, \hbar=\pi  \right) &= \frac{1}{2\sqrt{3}}, \\
Z\left(1, \hbar=2\pi \right) &= \frac{1}{9}, \\ 
Z\left(1, \hbar=4\pi \right) &= \frac{1}{36}.
\ea
\ee

This method of calculating the fermionic spectral traces becomes cumbersome when $N$ grows large. However, there is another, very efficient method to calculate the fermionic traces, first developed in \cite{hmo12a, py12}. This method is based on a set of integral TBA-like equations for the Fredholm determinant, which were first proposed in \cite{tw95} for a class of operators which includes some of the three-term operators \eqref{tto}. This Tracy--Widom method was later generalized in \cite{oz15} to \textit{all} operators of the form \eqref{tto}. Using these techniques, one can systematically compute the fermionic spectral traces recursively. For example, for $N=2$ one finds \cite{oz15}, 
\be
\label{N=2,hbar=1,2,4pi}
\ba
Z\left(2, \hbar=\pi  \right) &= \frac{1}{36}, \\
Z\left(2, \hbar=2\pi \right) &= -\frac{1}{81}+\frac{1}{12\sqrt{3}\pi}, \\
Z\left(2, \hbar=4\pi \right) &= \frac{5}{324}-\frac{1}{12\sqrt{3}\pi}.
\ea
\ee

As we mentioned above, one can use the conjecture put forward in \cite{ghm14a} to further compute the fermionic spectral traces numerically, for {\it arbitrary} values of $N$, in terms of BPS invariants of the CY $X$ (see \cite{ghm14a,km15,cgm15,oz15} for examples of this procedure, and \cite{hmo12b, hmo13, cgm14} for similar examples in ABJM theory). In this paper, we will assume the validity\footnote{Note that even if the conjecture of \cite{ghm14a} turned out to be false as an exact formula for the fermionic spectral traces, it is certainly valid with a precision which is much higher than the precision achieved by our resurgent analysis. For example, if we calculate $Z(2,4\pi)$ for local $\IP^2$ by using this conjecture with BPS invariants up to degree $7$, we match the exact, known value obtained in spectral theory with a precision of 150 digits, which is far larger than the precision we have for the Borel--Pad\'e resummation.} of this conjecture, and we will use it to compute fermionic spectral traces numerically whenever an exact expression from spectral theory is lacking, as well as for non-integer values of $N$.

\section{Transseries Constructions for Topological Strings}\label{sec:transseries}

Let us now shift gears and address the (resurgent) transseries completion of the topological-string free energy, as constructed in \cite{cesv13, cesv14}. This completion is based upon a nonperturbative extension of the holomorphic anomaly equations of \cite{bcov93}, and all its perturbative and nonperturbative ingredients are explicitly calculable. We shall review this construction in the following, as it applies to the local $\BP^2$ CY geometry \cite{cesv14} (the general construction can be found in \cite{cesv13}).

Recall from \eqref{dP} that local $\BP^2$ is a toric geometry that can be defined as the total space of the bundle $\CO(-3) \to \BP^2$. It is customary to parametrize the moduli 
space of complex structures of its mirror curve by $z$, which was introduced in (\ref{z-def}). The topological-string free energies we shall describe next depend on both $z$ and its complex conjugate. The mirror map is given by
\begin{equation}
-t = \log z - 6z + 45 z^2 - 560 z^3 + \cdots.
\end{equation}
\noindent
The moduli space has three special points: large-radius at $z=0$, conifold at $z=-1/27$, and orbifold at $z=\infty$. However, for the purpose of resurgent analysis, it was shown in \cite{cesv14} that the appropriate coordinate is $\psi$, related to $z$ by $z = (-3\psi)^{-3}$. The cubic root splits the conifold point into three values of $\psi$, at the cubic roots of the identity. These three conifold points---alongside the large-radius point---are each associated to an instanton action and, for this work, they are the relevant points in moduli space.


\subsection{General Transseries and Resurgence Relations}

The nonperturbative free energy $F_{\BP^2}$ has a transseries representation where $g_{\text{s}}$ is the resurgent variable. We shall shortly explain what the terms \textit{transseries} and \textit{resurgence} mean (full details on the results of this section can be found in \cite{cesv13, cesv14}), but let us proceed step-by-step.

The starting point is the perturbative series 
\begin{equation}
\Fn{0} (g_{\text{s}};z,\bar{z}) \simeq \sum_{g=0}^{+\infty} g_{\text{s}}^{2g-2}\, \Fng{0}{g} (z,\bar{z}).
\end{equation}
\noindent
Here the string coupling $g_{\text{s}}$ is regarded as a small formal parameter. The series on the right-hand side does not converge: it is \textit{asymptotic}. The reason is that its coefficients, \textit{i.e.}, the genus-$g$ free energies, grow faster than exponential: they grow factorially as $(2g)!$ when $g\to+\infty$ (technically this makes $\Fn{0} (g_{\text{s}})$ a Gevrey-1 series). The standard procedure tries to bypass this state of affairs by considering the Borel transform of $\Fn{0} (g_{\text{s}};z,\bar{z})$, a new series with a non-zero radius of convergence on the Borel $s$-plane, given by
\begin{equation}
\CB [\Fn{0}] (s) := \sum_{g=0}^{+\infty} \frac{\Fng{0}{g} (z,\bar{z})}{(2g)!}\, s^{2g-2}.
\end{equation}
\noindent
If the Borel transform of $\Fn{0}$ only has isolated singularities, the theory of resurgence can handle series of this kind and be used to properly define a function out of an initially asymptotic series. In this case $\Fn{0}$ is called a resurgent function, and the whole machinery of resurgence can begin its motion (we refer the reader to \cite{abs16} for an introduction and a complete list of references).

\begin{figure}[t!]
\begin{center}
\includegraphics[width=0.5\textwidth]{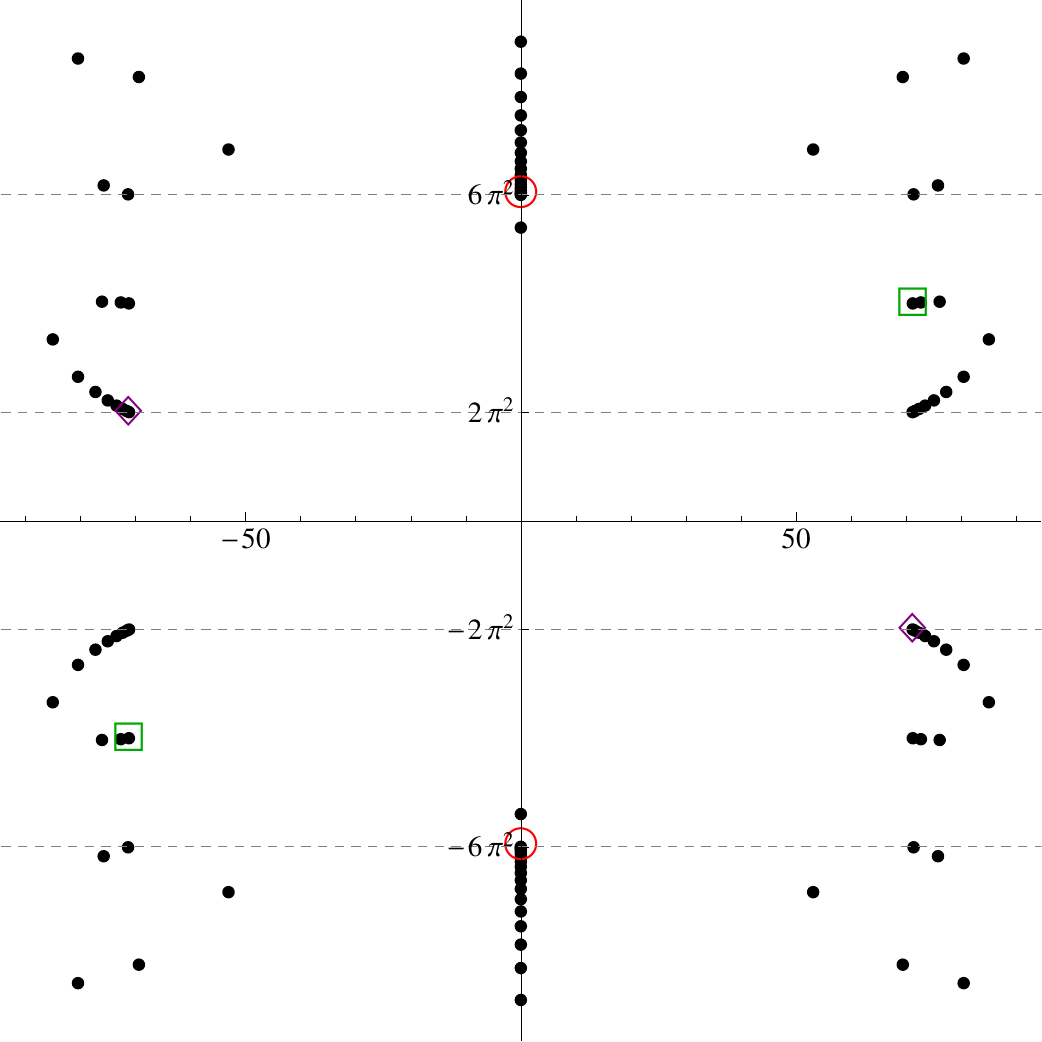}
\end{center}
\caption{The singularities of the Borel transform give us information about the instanton actions of the transseries. Herein we plot the poles of the Pad\'e approximant to the Borel transform of $\Fn{0}$, for $\lambda = \frac{3}{4\pi}$ (see \eqref{lambda,hbar,lambda}). The red circle depicts the first conifold instanton-action, $A_1$; the green square the second conifold instanton-action, $A_2$; and the purple diamond the large-radius instanton-action, $A_\txK$. It should be somewhat clear from the plot how these are branch points for the branch cuts of $\CB [\Fn{0}]$, with poles organizing as a leading singularity followed by a tail that represent the branch cut. All these quantities will be explained later in the main text.}
\label{fig:borel_plane_1}
\end{figure}

In the case of topological string theory, we have no rigorous proof that $\Fn{0}$ is resurgent but numerical evidence is abundant. In the example of local $\BP^2$, \figref{fig:borel_plane_1} shows a plot of the Borel plane for a particular value\footnote{We shall see in a moment that this plot does not depend on $\bar{z}$.} of $z$ and with the singularities of $\CB[\Fn{0}]$ highlighted. The first prediction of resurgence is that singularities on the Borel plane correspond to (nonperturbative) instanton actions of the full free energy $\FPtwo$, or at least of the formal transseries that represents $\FPtwo$. These instanton actions are special components of the transseries for $\FPtwo$, where in the present context a transseries is a formal series in both $g_{\text{s}}$ and $\rme^{-A/g_{\text{s}}}$, with $A$ one of the instanton actions (note how the exponential contribution is non-analytic at $g_{\text{s}}=0$, hence cannot be expanded). The simplest transseries, with just one instanton action, would be of the form
\begin{equation}
\label{eq:simplest_transseries}
\sum_{g=0}^{+\infty} g_{\text{s}}^{2g-2}\, \Fng{0}{g} + \sum_{n=1}^{+\infty} \sigma^n\, \rme^{-n A/g_{\text{s}}}\, g_{\text{s}}^{\stpwn{n}}\, \sum_{g=0}^{+\infty} g_{\text{s}}^g\, \Fng{n}{g}.
\end{equation}
\noindent
Here $\sigma$ is the transseries parameter, a complex number that keeps track of the instanton number $n$, but also encodes any freedom in initial or boundary conditions. Its value will be very important for the resummation in section~\ref{sec:resummation}. The characteristic exponent $\stpwn{n}$ is simply the starting power of $g_{\text{s}}$. As to the new coefficients, $\Fng{n}{g}(z,\bar{z})$, they encode the nonperturbative effects around each instanton sector, labeled by $(n)$ and associated to a total instanton action $nA(z,\bar{z})$.

For each $n$ we expect the series $\sum g_{\text{s}}^g\, \Fng{n}{g}$ to be asymptotic, with coefficients $\Fng{n}{g}$ growing as $\sim g!$. Altogether, \eqref{eq:simplest_transseries} forms a \textit{resurgent} transseries which extends $\Fn{0} (g_{\text{s}})$ and, in principle, captures all nonperturbative features\footnote{At least the ones associated to D-branes. There are also NS-brane effects \cite{c15, ayz15}, which are not needed herein.} of $F$. We shall later see how this is also true in practice. The qualifier ``resurgent'' implies that there is a precise relation between the coefficients $\Fng{n}{g}$ for different values of $n$ and $g$. The derivation of these relations can be found elsewhere (again, we refer the reader to \cite{abs16} for details on resurgent transseries), but the simplest of them would be
\be
\label{eq:simplest_resurgence_relations}
\ba
\Fng{0}{g} &\simeq \frac{\Gamma(2g-1)}{A^{2g-1}}\, \frac{S_1}{2\pi\rmi}\, \Fng{1}{0} + \frac{\Gamma(2g-2)}{A^{2g-2}}\, \frac{S_1}{2\pi\rmi}\, \Fng{1}{1} + \cdots + \\
&+ \frac{\Gamma(2g-1)}{(2A)^{2g-1}}\, \frac{S_1^2}{2\pi\rmi}\, \Fng{2}{0} + \frac{\Gamma(2g-2)}{(2A)^{2g-2}}\, \frac{S_1^2}{2\pi\rmi}\, \Fng{2}{1} + \cdots + \\
& + \cdots, \qquad \text{as $g\to +\infty$.}
\ea
\ee
\noindent
This means that, codified in the details of the factorial growth of $\Fng{0}{g}$, are all other (nonperturbative) coefficients $\Fng{n}{h}$, alongside the instanton action $A$. In this expression $S_1$ denotes the Stokes constant for the problem, to play a pivotal role in our resummations of section~\ref{sec:resummation}.

\subsection{Perturbative and Nonperturbative Sectors of Local $\BP^2$}

Resurgence relations such as \eqref{eq:simplest_resurgence_relations}, among others, are of general nature and can be used to discover and compute nonperturbative quantities, such as $A$ and $\Fng{n}{h}$, whenever only perturbative data is available. This is, in part, what was done in \cite{cesv14}: in that work the large-order (large $g$) relations \eqref{eq:simplest_resurgence_relations} were used, alongside a nonperturbative extension of the holomorphic anomaly equations \cite{cesv13}, to uncover and compute the transseries for the topological-string free energy on local $\BP^2$.

Let us be more detailed on these results. The holomorphic anomaly equations of \cite{bcov93} describe the nonholomorphic dependence of $\Fng{0}{g} (z,\bar{z})$, in terms of $\Fng{0}{h} (z,\bar{z})$ for $h=1,\ldots,g-1$. This recursiveness allows us to solve for $\Fng{0}{g}$ up to a function of $z$ alone, called the holomorphic ambiguity. For local $\BP^2$ (and many other examples) this ambiguity may be fixed by looking at the behavior of $\Fng{0}{g}$ at the large-radius and conifold points in moduli space, see, \textit{e.g.}, \cite{bcov93, hk06, gkmw07, hkr08, cesv13, cesv14}. The final result can be written down in quite a compact form, if we use the so-called propagator variable\footnote{The reader should not confuse propagator variable and Stokes constants, in spite of similar notation.} $\Szz(z,\bar{z})$ which takes the place of $\bar{z}$. In this way \cite{yy04, al07},
\begin{equation}
\Fng{0}{g} = \Pol(\Szz; 3g-3) \equiv \text{polynomial in $\Szz$ of degree $3g-3$ with rational coefficients in $z$}.
\end{equation}
\noindent
For example, the propagator is expressed in terms of $\Fng{0}{1}$ through
\begin{equation}
\p_z \Fng{0}{1} = \frac{1}{2} C_{zzz} \Szz,
\end{equation}
\noindent
where $C_{zzz} = (-3 z^3(1+27z))^{-1}$ is the Yukawa coupling for local $\BP^2$. The genus $g=2$ free energy then satisfies the equation
\begin{equation}
\p_{\Szz} \Fng{0}{2} = \frac{1}{2} \left( D_z \p_z \Fng{0}{1} + \left(\p_z \Fng{0}{1}\right)^2 \right),
\end{equation}
\noindent
which can be integrated with respect to $\Szz$, and its ambiguity fixed, to 
\begin{equation}
\Fng{0}{2} = C_{zzz}^2 \left( \frac{5}{24} (\Szz)^3 - \frac{3z^2}{16} (\Szz)^2 + \frac{z^4}{16} \Szz - \frac{(11 - 162z-729z^2)z^6}{1920} \right).
\end{equation}
\noindent
See \cite{hkr08} for details on computing the perturbative series and fixing its holomorphic ambiguities.

There are analogous equations for the $\Fng{n}{g}$ that can be solved recursively in $n$ and $g$, as explained in general in \cite{cesv13}. These sectors also have their holomorphic ambiguities, which may be determined by making use of equation \eqref{eq:simplest_resurgence_relations}. The idea is to take the holomorphic limit of \eqref{eq:simplest_resurgence_relations}, which will maintain the holomorphic ambiguities, then compare expansions in $1/g$ and thus solve for the unknown ambiguities. Note, however, that one cannot disentangle the ambiguity from the Stokes constant $S_1$; it is only possible to compute their product. Let us show this with an example. Let $n=1$ and $h=0$. The extended holomorphic anomaly equations imply \cite{cesv13}
\begin{equation}
\Fng{1}{0}(z, \Szz) = \fng{1}{0}(z)\, \rme^{\frac{1}{2} (\p_z A)^2 \Szz},
\end{equation}
\noindent
where $\fng{1}{0}(z)$ is the holomorphic ambiguity. Assume for example that $A$ is $A_\txK = -2 \pi t $, the K\"ahler instanton action. Now, on the one hand, we can take the leading term in equation \eqref{eq:simplest_resurgence_relations} in the holomorphic limit of the large-radius frame, for which\footnote{We follow the notation in \cite{cesv13, cesv14}: roman letters ($F$, $S^{zz}$) indicate nonholomorphic quantities, curly letters ($\CF$, $\CS^{zz}$) indicate their holomorphic limits. The chosen frame is shown as a subscript separated by a semicolon (;).}
\begin{equation}
\label{eq:Szz_hol_AK}
\hSzzA{A_\txK} = \frac{2}{C_{zzz}} \p_z \hFngA{0}{1}{A_\txK} = \frac{2}{C_{zzz}} \left( \frac{1}{12z\left(1+27z\right)} - \frac{{}_2F_1 \left. \left( \frac{2}{3}, \frac{4}{3}, 1\, \right| -27z \right)}{6z\, {}_2F_1 \left. \left( \frac{1}{3}, \frac{2}{3}, 1\, \right| -27z \right)} \right).
\end{equation}
\noindent
On the other hand, large-order numerical computations (see appendix~\ref{sec:checks_1instanton_AK}) show that
\begin{equation}
\label{eq:glimit_AK_loop_1}
\lim_{g\to+\infty} \frac{A_\txK^{2g-1}}{\Gamma(2g-1)}\, \hFngA{0}{g}{A_\txK} = 3\, \frac{A_\txK}{2\pi^2}.
\end{equation}
\noindent
Since the left-hand side of this equation is also $\frac{\stk{A_\txK}}{\rmi\pi}\, \hFnAgA{1}{A_\txK}{0}{A_\txK}$, we conclude\footnote{The notation $\instsect{n}{A}$ means that only the sector associated to the instanton action $A$ is turned on, at instanton number $n$, while all others are set to zero.}
\begin{equation}
\frac{\stk{A_\txK}}{\rmi\pi} \FnAg{1}{A_\txK}{0} = 3\, \frac{A_\txK}{2\pi^2}\, \rme^{\frac{1}{2}(\p_z A_\txK)^2 \left( \Szz - \hSzzA{A_\txK} \right)}.
\end{equation}

A similar strategy can be implemented for higher loops and other multi-instanton sectors. The results grow in complexity but we can give a brief description of the general picture \cite{cesv13, cesv14}:
\begin{itemize}
\item There are four relevant instanton actions: $A_1$, $A_2$, $A_3$, and $A_\txK$. They are all holomorphic (\textit{i.e.}, independent of $\Szz$) in accordance to arguments in \cite{dmp11} and the extended holomorphic anomaly equations of \cite{cesv13}. The first three actions are related to the conifold point at $z= -1/27$, or $\psi = 1$, $\rme^{+2\pi\rmi/3}$ and $\rme^{-2\pi\rmi/3}$, and they enjoy a $\BZ_3$-symmetry among themselves:
\begin{gather}
A_1(\psi) = \frac{2\pi\rmi}{\sqrt{3}}\, t_\txc(\psi), \qquad A_2(\psi) = A_1(\psi\,\rme^{-2\pi\rmi/3}), \qquad A_3(\psi) = A_1(\psi\,\rme^{+2\pi\rmi/3}), \\
t_\txc(\psi) = \frac{2\pi}{\sqrt{3}} \left( \frac{3\psi}{\G\left(\frac{2}{3}\right)^3}\,\, {}_3 F_2 \left. \left( \frac{1}{3},\frac{1}{3},\frac{1}{3}; \frac{2}{3},\frac{4}{3}\, \right| \psi^3 \right) - \frac{\frac{9}{2}\psi^2}{\G\left(\frac{1}{3}\right)^3}\,\, {}_3 F_2 \left. \left( \frac{2}{3},\frac{2}{3},\frac{2}{3}; \frac{4}{3},\frac{5}{3}\, \right| \psi^3 \right) - 1 \right),
\end{gather}
\noindent
where, as mentioned, $\psi = (-27z)^{-1/3}$. The last instanton action, $A_\txK$, we have already seen that it is proportional to the K\"ahler parameter,
\bea
A_\txK &=& -2 \pi t, \\
t(z) &=& \frac{\sqrt{3}}{2\pi}\, G^{22}_{33}{\left( \begin{array}{ccc|} 1/3 & 2/3 & 1 \\ 0 & 0 & 0 \end{array} \; 27z \right)}.
\eea
\noindent
On top of these four actions there is a trivial instanton action, $4\pi^2\rmi$, stemming from the constant map contribution to $\Fng{0}{g}$ \cite{bcov93, ps09, cesv13, cesv14}. It will play no role in what follows.
\item Each of the previous instanton actions $A$ comes paired with its opposite $-A$ (which is a consequence of the string genus expansion). There is, however, a symmetry between the corresponding sectors $\FnAg{n}{+A}{g}$ and $\FnAg{n}{-A}{g}$ \cite{gikm10, asv11} so that we only need to worry about the first of them. Further, for the particular values of $z$ which we shall consider, the Borel plane has a reflexive symmetry along the real axis, and there is a conjugation symmetry we should also take into account. 
\item Each instanton action leads to a nonperturbative sector, for which we have corresponding free energies, $\FnAg{n}{A}{g}(z,\Szz)$. For our purposes in this paper, it is enough to look at the first instanton sectors: $n_i=1$ for some $i$ and $n_j =0$ for $j\neq i$ (see \cite{cesv13, cesv14} for $n\geq2$ and their associated subtleties). The anti-holomorphic structure of $\FnAg{1}{A}{g}$ is
\begin{equation}
\FnAg{1}{A}{g} = \rme^{\frac{1}{2}(\p_z A)^2 \left( \Szz - \hSzzA{A} \right)}\, \Pol(\Szz; 3g),
\end{equation}
\noindent
where the coefficients involve polynomials in $A$, $\p_z A$, $\p_z^2 A$, and rational functions of $z$ \cite{cesv13, cesv14}. The holomorphic ambiguity is fixed with the holomorphic conditions
\begin{equation}
\label{eq:generic_holo_limit}
\frac{\stk{A}}{\rmi\pi}\, \hFnAgA{1}{A}{0}{A} = \alpha_A\, \frac{A}{2\pi^2}, \qquad \frac{\stk{A}}{\rmi\pi}\, \hFnAgA{1}{A}{1}{A} = \alpha_A\, \frac{1}{2\pi^2}, \qquad \frac{\stk{A}}{\rmi\pi}\, \hFnAgA{1}{A}{g\geq 2}{A} = 0.
\end{equation}
\noindent
Here we take the holomorphic limit in the frame associated to $A$ (conifold or large-radius). $\alpha_A = 1$ if $A$ is one of the three conifold actions, and $\alpha_A =3$ if $A$ is the K\"ahler action.
\item Note how each nonperturbative sector has its own Stokes constant $\stk{A}$, which will appear inverted $\stk{A}^{-1}$ as a pre-factor of $\FnAg{1}{A}{g}$. Therefore, in the trasseries we will always find it alongside the transseries parameter $\sgm{A}$, as $\frac{\rmi\pi\,\sgm{A}}{\stk{A}}$. This is the combined quantity which will have to be fixed\footnote{Note that it was argued in \cite{cesv14} that the Stokes constant should not depend on the moduli, because that would be incompatible with numerical checks of the resurgent relations.} when we resum the transseries in section~\ref{sec:resummation}.
\item The holomorphic anomaly equations do not fix the power of the exponent $\stpwnA{1}{A}$ in the transseries, but the resurgence results suggest that $\stpwnA{1}{A}=-1$ is the correct value.
\item Explicitly, the one-instanton free energies $\FnAg{1}{A}{g}$ for low $g$ are
\begin{align}
\frac{\stk{A}}{\rmi\pi}\, \FnAg{1}{A}{0} &= \frac{\alpha_A}{2\pi^2}\, \rme^{\frac{1}{2}a_1^2\, \Delta S}\, a_0, \\
\frac{\stk{A}}{\rmi\pi}\, \FnAg{1}{A}{1} &= \frac{\alpha_A}{2\pi^2}\, \rme^{\frac{1}{2}a_1^2\, \Delta S} \left\{ 1 - \frac{1}{2} \left( 2 a_1 + a_0 C \CS \right) a_1 \Delta S - \frac{1}{2} a_0 a_1 C (\Delta S)^2 - \frac{1}{6} a_0 a_1^3 C (\Delta S)^3 \right\}, \\
\frac{\stk{A}}{\rmi\pi}\, \FnAg{1}{A}{2} &= \frac{\alpha_A}{2\pi^2}\, \rme^{\frac{1}{2}a_1^2\, \Delta S}\, a_1^2\, C\, (\Delta S)^2 \left\{ \frac{1}{16} \left( 8 a_1 \CS + a_0 C \left( 10 \CS^2 - 6 \CS z^2 + z^4 \right) \right) + \right. \nonumber\\
& \left. + \frac{1}{8} \left( \frac{20}{3} a_1 + a_0 C \left( 10 \CS - 3 z^2 \right) \right) \Delta S + \right. \\
& \hspace{-40pt} \left. + \frac{1}{48} \left( 8 a_1^3 + a_0 C \left( 30 + a_1^2 \left( 10 \CS - 3 z^2 \right) \right) \right) (\Delta S)^2 + \frac{5}{24} a_0 a_1^2 C (\Delta S)^3 + \frac{1}{72} a_0 a_1^4 C (\Delta S)^4 \right\}, \nonumber
\end{align}
\noindent
where we have introduced the notation
\begin{alignat}{2}
&a_k := \partial_z^k A \qquad (k=0,1,2), \qquad &\Delta S := \Szz- \hSzzA{A}, \\
&C := C_{zzz} = -\frac{1}{3 z^3 (1+27z)}, \qquad &\CS := \hSzzA{A} = -\frac{1}{C_{zzz}} \frac{a_2}{a_1} + \frac{1}{2} z^2 (7+216z).
\end{alignat}
\noindent
There are no third or higher derivatives of $A$ because we used the (third order) Picard--Fuchs equation, and there are no second derivatives as we traded them by $\partial_z A$ and $\hSzzA{A}$.
\item In summary, the transseries is, to leading instanton order,
\begin{gather}
\sum_{g=0}^{+\infty} g_{\text{s}}^{2g-2}\, \Fng{0}{g} + \sum_{A\in\{A_1,A_2,A_3,A_\txK\}} \sgm{A}\, \rme^{- A/g_{\text{s}}}\, g_{\text{s}}^{-1} \sum_{g=0}^{+\infty} g_{\text{s}}^g\, \FnAg{1}{A}{g} + \nonumber \\
+ \text{ (opposite actions) } + \text{ (complex conjugate actions) }.
\label{eq:transseries_to_first_order}
\end{gather}
\end{itemize}

The goal now is to resum this object, after having fixed the transseries parameters, and compare against the results of section~\ref{sec:exact}. In order to do that we must first make sure that all conventions agree, and, in particular, also need to decide which holomorphic limit must be taken. 

Let us address this second question. Recall from section~\ref{sec:exact} that the 't~Hooft limit \eqref{thooft-lam}, leading to the topological-string genus expansion, has 't~Hooft parameter given by \eqref{saddle} and associated to the conifold frame. In particular, for our example of local $\BP^2$, the 't~Hooft coupling $\lambda$ is proportional to the conifold flat coordinate $t_{\text{c}}$ via \eqref{'tHooft-tc}, which selects the \textit{first conifold frame}, associated to $A_1$, as the relevant one to use in the transseries. This implies that in the next section, when we resum each asymptotic series in \eqref{eq:transseries_to_first_order}, we must evaluate the propagator $\Szz$ to\footnote{Here, $P_\nu(x)$ and $Q_\nu(x)$ are Legendre functions of first and second kind, respectively.} 
\begin{equation}
\hSzzA{A_1} = \frac{z^2}{2} \left( - 1 - 54 z + 2 \frac{\pi\, P_{2/3} (1+54z) + 2\sqrt{3}\, Q_{2/3} (1+54z)}{\pi\, P_{-1/3} (1+54z) + 2\sqrt{3}\, Q_{-1/3} (1+54z)} \right).
\end{equation}
\noindent
Further recall that the dictionary between $(N,\hbar)$ and $(g_{\text{s}},z)$ is 
\begin{equation}
\label{lambda,hbar,lambda}
\lambda = \frac{N}{\hbar}, \qquad \hbar = \frac{4\pi^2}{g_{\text{s}}}, \qquad \lambda = \frac{t_c(z)}{4\sqrt{3}\pi^2}.
\end{equation}
\noindent
Finally, we must use
\begin{equation}
\label{eq:precise_perturbative_expansion}
\hFnfA{0}{A_1} = g_{\text{s}}^{-2} \left( \hFngA{0}{0}{A_1} - \frac{2\pi^2\rmi}{3\sqrt{3}}\, t_{\text{c}} \right) + \hFngA{0}{1}{A_1} - \frac{\log \hbar}{12} + \sum_{g=2}^{+\infty} g_{\text{s}}^{2g-2} \left( \hFngA{0}{g}{A_1} - \frac{(-1)^{g-1}B_{2g-2}B_{2g}}{3^{2g-2} 4g \left(2g-2\right) \left(2g-2\right)!}\right)
\end{equation}
\noindent
for the perturbative sector of the transseries, in order to connect with the free energies \eqref{gen-J-as}, \eqref{small-lam}, as obtained in spectral theory. From now on, we drop the subindex indicating the frame in which the holomorphic limit is taken, writing $\hFn{0}$ for $\hFnfA{0}{A_1}$, $\hFnAg{n}{A}{g}$ for $\hFnAgA{n}{A}{g}{A_1}$, and so on.

\section{Resummations and Agreement with the Exact Free Energy}\label{sec:resummation}

An asymptotic series (or even a transseries) by itself cannot yield numbers as it does not converge anywhere. A resummation procedure is thus required, and by this we mean an operation which takes such formal objects and out spits some honest function that we can evaluate numerically. In the resurgence context, the natural resummation method is that of Borel resummation for asymptotic series, and Borel--\'Ecalle resummation for transseries (where the latter is built from the former). Pad\'e approximants will be required to implement these resummations numerically.

Borel resummation is the composition of a Borel and a Laplace transform. Term-by-term in the series these two operations are inverses of each other, but on the analytic continuation of the Borel transform, the Laplace transform produces a function:
\bea
\sum_{g=0}^{+\infty} g_{\text{s}}^g\, \Fng{n}{g} &\longmapsto& \CB[\Fn{n}](s) = \sum_{g=0}^{+\infty} \frac{\Fng{n}{g}}{g!}\, s^g, \\
\CB[\Fn{n}](s) &\longmapsto& \CS_\theta \Fn{n}(g_{\text{s}}) = g_{\text{s}}^{-1}\, \int_0^{\rme^{\rmi\theta} \cdot \infty} \rmd s\, \CB[\Fn{n}](s)\, \rme^{-s/g_{\text{s}}}.
\eea
\noindent
When $g_{\text{s}}$ is real we take $\theta=0$ and integrate along the positive real axis on the Borel plane. This is valid as long as $\CB[\Fn{n}](s)$ has no real positive poles, or, in a different language, no instanton action has both a vanishing imaginary part and a positive real part. When this happens we are on a Stokes line and the resummation is ambiguous, depending on how to avoid the singularity. The difference between resummations above and below a Stokes line is of order $\rme^{-A/g_{\text{s}}}$, where $A$ is the singularity on the positive axis with the least real part. However, by considering the resummation at the transseries level, these nonperturbative differences simply interpolate between different nonperturbative sectors of the transseries and, in essence, give rise to the familiar Stokes phenomena. Borel resummation at the transseries level, which naturally incorporates Stokes phenomena, is dubbed Borel--\'Ecalle resummation, and some examples within the large $N$ context were considered in \cite{m08, csv15} (our strategy of calculation in this section follows these references). For the resummation of local $\BP^2$ we do not have enough precision to include sectors beyond $n=1$, but we will encounter a Stokes line that needs to be addressed in detail.

On a practical note, Borel resummation is approximated numerically with Borel--Pad\'e resummation. The idea is to use a Pad\'e approximant to the truncated Borel transform (there is only finite data). The poles of the Pad\'e approximant mimic the singularities of the real function in the same way we have used them to plot the instanton actions on the Borel plane in \figref{fig:borel_plane_1}. 

\subsection{Resummation of the Perturbative Free Energy}

Let us begin by addressing the resummation of the perturbative sector. It is natural to expect this sector to yield the leading contribution to the nonperturbative free energy, with (exponentially suppressed) instanton sectors in the transseries being subleading. As an example, consider $N=2$ and $\hbar = 4\pi$ for which
\be
\hFPtwo = \log \left( \frac{5}{324} - \frac{1}{12\sqrt{3}\pi} \right) = - \underline{9.049\,862\,10}2\,738\,02\ldots,
\ee
\noindent
as computed in \eqref{N=2,hbar=1,2,4pi}. The Borel--Pad\'e resummation\footnote{In practice, we have used data from \cite{cesv14}, with coefficients up to genus $g=114$.} of $\hFn{0}(g_{\text{s}})$ at the corresponding values of $z$ and $\hSzzA{A_1}$ yields, in turn,
\be
\CS_0 \hFn{0} = -\underline{9.049\,862\,10}3\,051\,21\ldots,
\ee
\noindent
where all displayed digits are stable. The results are rather similar, albeit not equal. The difference is small, of order $10^{-10}$, and if our reasoning holds it should be associated to a one-instanton effect, with weight $\re A \approx 70$ for some instanton action $A$ we have not identified yet (we shall soon see how to precisely match this effect with the one-instanton sector of the transseries).

\begin{figure}[t!]
\begin{center}
\includegraphics[width=0.8\textwidth]{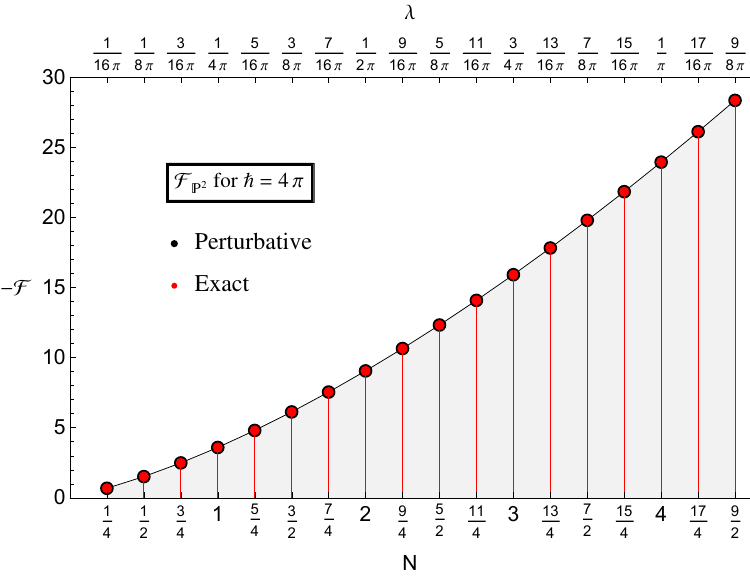}
\end{center}
\caption{Comparison of the perturbative resumation $\CS_0 \hFn{0}$ against the exact value $\hFPtwo$, for $\hbar=4\pi$ and several values of $N$, not necessarily integer. The small difference between the two cannot be appreciated with the naked eye.}
\label{fig:exact_vs_perturbative}
\end{figure}

The comparison between $\hFPtwo$ and $\CS_0 \hFn{0}$ may be done for many other values of $N$ and $\hbar$, or, equivalently, of $g_{\text{s}}$ and $z$. We show some such examples in \figref{fig:exact_vs_perturbative} and Table~\ref{tab:exact_vs_perturbative}.  \figref{fig:exact_vs_perturbative} plots values of $N$ at fixed $\hbar=4\pi$, and shows how close the exact and perturbative free energies are (visually they seem to agree but that is only because their difference is exponentially small). In Table~\ref{tab:exact_vs_perturbative} we picked varying values of $(N,\hbar)$, where one may appreciate how small the difference between $\hFPtwo$ and $\CS_0 \hFn{0}$ can be. This is expected, since, according to the conjecture in \cite{ghm14a,mz15}, the difference between the nonperturbative free energy $\hFPtwo$ and the Borel resummation of the perturbative series must be an exponentially small effect. In spite of this, it is important to stress that perturbation theory alone is simply \textit{not enough} to recover the full nonperturbative result $\hFPtwo$ of section~\ref{sec:exact}. Instead, one needs the transseries with its nonperturbative instanton contributions, to which we now turn. We should note that the situation encountered here is very similar to the one found for the $1/N$ expansion of the free energy of ABJM theory on the three-sphere: this expansion is Borel summable, for generic values of the 't~Hooft parameter \cite{dmp10,dmp11}, but its Borel resummation does not agree with the exact value \cite{gmz14}. The difference between these quantities is due to a large $N$ instanton effect. 

\begin{table}[t!]
\centering
\ra{1.3}
\begin{tabular}{llll}
\toprule
\multicolumn{1}{c}{$N$} & \multicolumn{1}{c}{$\hbar$} & \multicolumn{1}{c}{$\hFPtwo$} & \multicolumn{1}{c}{$\CS_0 \hFn{0}$} \\
\midrule
$\frac{1}{2}$ & $4 \pi$ & $-\underline{1.505\,134}\,688\,359$ &  $-\underline{1.505\,134}\,801$ \\
$4$ & $\frac{7\pi}{4}$ & $-\underline{14.769\,466}\,431\,898$ & $-\underline{14.769\,466}\,390$ \\
$3$ & $2\pi$ & $-\underline{10.476\,827}\,230\,709$ & $-\underline{10.476\,827}\,144$ \\
$\frac{3}{2}$ & $3\pi$ & $-\underline{5.062\,186\,2}71\,428$ & $-\underline{5.062\,186\,2}26\,8$ \\
$2$& $4 \pi$ & $-\underline{9.049\,862\,10}2\,738$ & $-\underline{9.049\,862\,10}3\,051$ \\
$\frac{1}{2}$ & $6\pi$ & $-\underline{2.059\,410\,649}\,773$ & $-\underline{2.059\,410\,649}\,10$ \\
$\frac{5}{2}$ & $3\pi$ & $-\underline{10.324\,978\,18}6\,430$ & $-\underline{10.324\,978\,18}3\,76$ \\
$3$ & $\frac{24\pi}{5}$ & $-\underline{17.784\,899\,587\,4}49\,32$ & $-\underline{17.784\,899\,587\,4}50\,79$ \\
\bottomrule
\end{tabular}
\caption{Comparison of exact $\hFPtwo$ and perturbative resummation $\CS_0 \hFn{0}$, for several values of $N$ and $\hbar$. Common digits are underlined (all displayed digits are numerically stable).}
\label{tab:exact_vs_perturbative}
\end{table}

\subsection{Resummation of the One-Instanton Sector}

Recall from section~\ref{sec:transseries} that the transseries for local $\BP^2$ has several instanton actions, with their corresponding nonperturbative sectors. To leading order, the small difference between $\hFPtwo$ and $\CS_0 \hFn{0}$ that we have just discussed, should be a one-instanton effect, \textit{i.e.}, of order $\rme^{-A/g_{\text{s}}}$. But this could hold for one or more of the possible instanton actions $A_1$, $A_2$, $A_3$, $A_\txK$ (along with their negatives and complex conjugates). There are some constraints, however. Since the difference $\hFPtwo - \CS_0 \hFn{0}$ is small, we should look for $A$ with $\re A>0$, and since it is real we should include $\bar{A}$ as well. Now, for $\lambda>0$ the action $A_1$ is purely imaginary and we can discard it from the resummation, since $\rme^{-A_1/g_{\text{s}}}$ would be a pure phase\footnote{One can use the same argument to discard the action $4\pi^2\rmi$, coming from the constant-map contribution.} and not exponentially suppressed. Out of the remaining three actions, $A_2$ is the natural choice in view of the computation we shall describe next (see appendix \ref{sec:relations_sectors} for comments on the possible roles of the sectors related to $A_3$ and $A_\txK$).

Let us thus consider the $A_2$-sector alongside its complex conjugate. Note that the explicit function $A_2$ has a branch cut along $\lambda >0$, and we choose the convention of evaluating $A_2(\lambda + \rmi\epsilon)$ as $\epsilon\to 0^+$. The goal of this section is then to match the reported difference $\hFPtwo - \CS_0 \hFn{0}$, with the one-instanton contributions arising from $A_2$ and $\widebar{A_2}$, \textit{i.e.}, to match
\begin{equation}
\label{eq:diff_match_1instA2_v1}
\hFPtwo - \CS_0 \hFn{0} = \sgm{A_2}\, g_{\text{s}}^{-1}\, \rme^{-A_2/g_{\text{s}}}\, \CS_0 \hFnA{1}{A_2} + \text{ (c.c.)},
\end{equation}
\noindent
where (c.c.) indicates complex conjugate\footnote{Note that here $z$ and $g_{\text{s}}$ are real, in which case taking the complex conjugate of the whole instanton contribution for $A_2$ is equivalent to substituting $A_2 \to \widebar{A_2}$.}. When implementing the calculation, it is important to recall that $\CS_0 \hFnA{1}{A_2}$ indicates the Borel(--Pad\'e) resummation of the holomorphic limit of the free energy, \textit{in the conifold-1 frame} (recall the last paragraph of section~\ref{sec:transseries}). In practice we perform the resummation with 16 coefficients of $\hFnAg{1}{A_2}{g}$, using data from \cite{cesv14}, which proves to be enough for the match at leading order\footnote{Working in the holomorphic limit of the frame associated to $A_1$, implies that  $\hFnAfA{1}{A}{A_1} (g_{\text{s}}) \simeq \sum_{g=0}^{+\infty} g_{\text{s}}^g\,\hFnAgA{1}{A}{g}{A_1}$ only truncates when $A = A_1$ (via \eqref{eq:generic_holo_limit}). For all other sectors we have an infinite asymptotic series \cite{cesv14}.}. Checking higher instanton corrections would require higher precision and many more coefficients (which is computationally intricate).

At this stage, the only unknown in equation \eqref{eq:diff_match_1instA2_v1} is $\sgm{A_2}$. Fixing the transseries parameter is, however, a bit more subtle in the topological string context, than it was in the matrix model context of \cite{m08, csv15}. In fact, due to the way in which the nonperturbative holomorphic ambiguities were fixed, we only have access to the combinations $\frac{\stk{A_2}}{\rmi\pi}\, \hFnAg{1}{A_2}{g}$. It is thus convenient to write the right-hand side of equation \eqref{eq:diff_match_1instA2_v1} as
\begin{equation}
\text{right-hand side of \eqref{eq:diff_match_1instA2_v1}} = \frac{\rmi\pi\, \sgm{A_2}}{\stk{A_2}}\, g_{\text{s}}^{-1}\, \rme^{-A_2/g_{\text{s}}}\, \CS_0 \left( \frac{S_{A_2,1}}{\rmi\pi}\, \hFnA{1}{A_2} \right) + \text{ (c.c.)}.
\end{equation}
\noindent
If we define the ``reduced'' transseries parameter and the ``reduced'' free energies as
\begin{equation}
\tsgm2 := \frac{2\pi\rmi}{\stk{A_2}}\, \sgm{A_2}, \qquad \thFnA{1}{A_2} :=  \frac{S_{A_2,1}}{2\pi\rmi}\, \hFnA{1}{A_2},
\end{equation}
\noindent
we obtain the final and useful form of equation \eqref{eq:diff_match_1instA2_v1},
\begin{equation}
\label{eq:diff_match_1instA2_v2}
\hFPtwo - \CS_0 \hFn{0} = \re \left( \tsgm2\, g_{\text{s}}^{-1}\, \rme^{-A_2/g_{\text{s}}}\, \CS_0 \thFnA{1}{A_2} \right) \equiv \CS_0 \Phi^{(1)}.
\end{equation}
\noindent
In this final form, our check will be achieved once we find a value for the ``reduced'' transseries parameter $\tsgm2$, such that this equation holds for all values of $(N,\hbar)$ we shall be considering. 

Let us immediately state our final result, and then build up to it. After looking at several points in the $(N,\hbar)$-plane we find that there are two distinct regions on this plane, on which $\tsgm2$ takes different values. Their border is found at 
\begin{equation}
\lambda = \frac{N}{\hbar} = \frac{1}{4\pi},
\end{equation}
\noindent
which is a straight line on the $(N,\hbar)$-plane, going through the origin. The conjectured exact value of $\tsgm2$ on both sides is
\begin{equation}
\label{eq:conjectured_tilde_sigma2}
\tsgm2 = \begin{cases}
2\pi\rmi\, \rme^{2\pi\rmi N}, & \lambda > \frac{1}{4\pi}, \\
2\pi\rmi \left(\rme^{2\pi\rmi N}-1\right), & \lambda < \frac{1}{4\pi}.
\end{cases}
\end{equation}
\noindent
This has the full flavor of Stokes phenomenon, but before giving a definitive interpretation of this result let us discuss the numerical support for equation \eqref{eq:conjectured_tilde_sigma2}.

\begin{table}[t!]
\centering
\ra{1.3}
\begin{tabular}{lllll}
\toprule
\multicolumn{1}{c}{$N$} & \multicolumn{1}{c}{$\hbar$} & \multicolumn{1}{c}{$\hFPtwo$} & \multicolumn{1}{c}{$\CS_0 \hFn{0}$} & \multicolumn{1}{c}{$\CS_0 \hFn{0} + \CS_0 \Phi^{(1)}$} \\
\midrule
$\frac{3}{2}$ & $2\pi$ & $-\dblunderline{3.862\,59}{2\,40}4\,378\,576\,04$ & $-\underline{3.862\,59}0\,319$ & $-\uwave{3.862\,592\,40}$ \\
$3$ & $2\pi$ & $-\dblunderline{10.476\,827}{\,230}\,709\,007\,0$ & $-\underline{10.476\,827}\,144$ & $-\uwave{10.476\,827\,230}$ \\
$\frac{9}{4}$ & $3\pi$ & $-\dblunderline{8.896\,384\,3}{90\,454}\,244\,69$ & $-\underline{8.896\,384\,3}85\,261$ & $-\uwave{8.896\,384\,390\,454}$ \\
$2$ & $4\pi$ & $-\dblunderline{9.049\,862\,10}{2\,738\,02}0\,42$ & $-\underline{9.049\,862\,10}3\,051\,20$ & $-\uwave{9.049\,862\,102\,738\,02}$ \\
\bottomrule  
\end{tabular}
\caption{Comparison of the exact $\hFPtwo$, the perturbative resummation, and the perturbative plus one-instanton resummation, for several values of $N$ and $\hbar$ with $\lambda > \frac{1}{4\pi}$. The value of $\tsgm2$ is that in \eqref{eq:conjectured_tilde_sigma2}. Common digits are marked with straight (perturbative) and wavy (perturbative plus one-instanton) underlines (all displayed digits are stable).}
\label{tab:exact_vs_perturbative_vs_oneinstanton_right}
\end{table}

In Table~\ref{tab:exact_vs_perturbative_vs_oneinstanton_right} we display, for several values of $(N,\hbar)$ with $\lambda > \frac{1}{4\pi}$, the exact free energy, $\hFPtwo$, the perturbative resummation, $\CS_0 \hFn{0}$, and the sum of perturbative and one-instanton contributions, $\CS_0 \hFn{0} + \CS_0 \Phi^{(1)}$. It is clear that using the value of $\tsgm2$ in \eqref{eq:conjectured_tilde_sigma2} we match the exact result for the nonperturbative free energy, and the transseries resummation, with increasing accuracy.

\begin{figure}[t!]
\begin{center}
\includegraphics[width=1\textwidth]{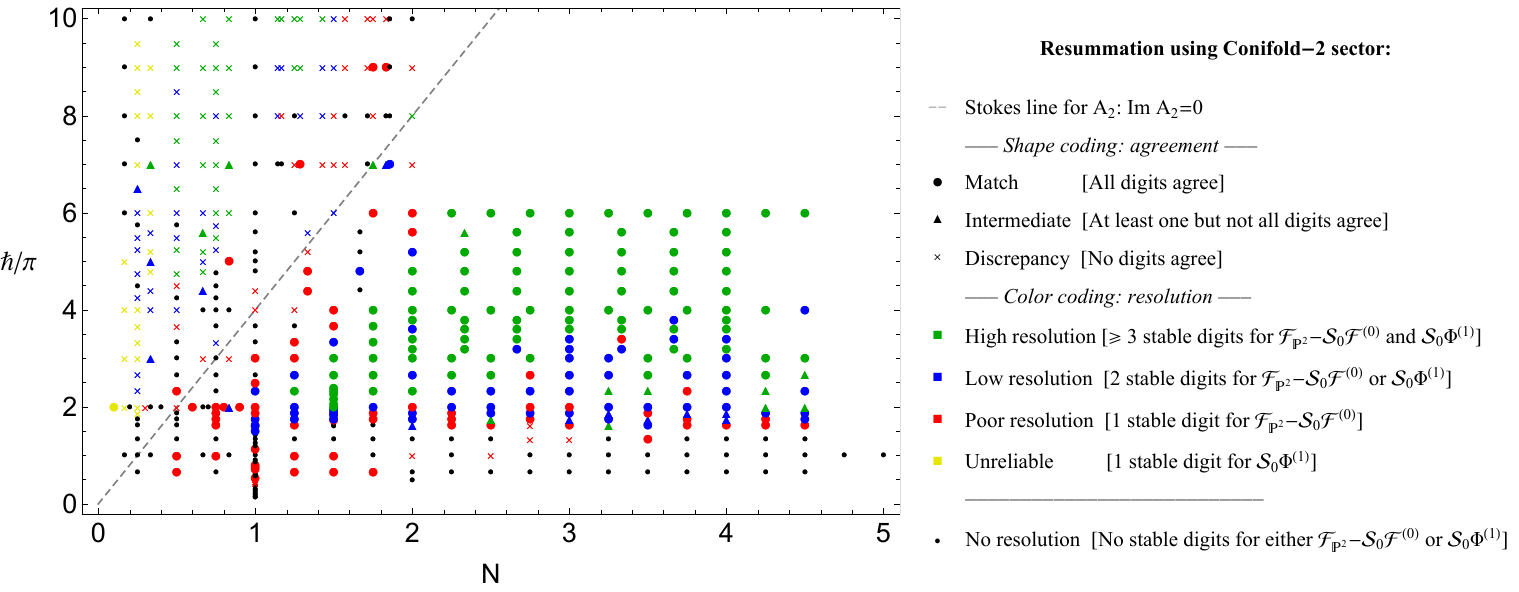}
\end{center}
\caption{This plot shows the agreement between $\hFPtwo- \CS_0 \hFn{0}$ and the one-instanton contribution in the right-hand side of equation \eqref{eq:diff_match_1instA2_v2}. We use $\tsgm2 = 2\pi\rmi\, \rme^{2\pi\rmi N}$, as in \eqref{eq:conjectured_tilde_sigma2} for $\lambda > \frac{1}{4\pi} $. The coding is as follows. Whenever there is complete agreement we indicate it with a circle. The triangle indicates disagreement in some of the digits, and the cross means total disagreement. The reliability of comparison may also vary. The color green means that at least 3 stable digits are available for comparison. The color blue indicates 2 stable digits for either quantity, and red means only 1 stable digit for $\hFPtwo - \CS_0 \hFn{0}$. Yellow is the least reliable color: only 1 stable digit for the one-instanton contribution. The dashed line is the Stokes line, given by $\im A_2 = 0$.}
\label{fig:agreement_N_hbar}
\end{figure}

This computation can be automated and repeated for many other values of $(N,\hbar)$, as depicted in \figref{fig:agreement_N_hbar}. The symbolic coding in the figure transmits how precise is the match between $\hFPtwo- \CS_0 \hFn{0}$ and $\CS_0 \Phi^{(1)}$. A circle indicates a perfect match, while complete disagreement is marked with a cross. Partial agreement (meaning only some digits) is signaled with a triangle. The colors of these symbols then indicate the number of available (stable) digits to match. Green denotes at least three stable digits, blue means two, while red and yellow mean just one. When no stable digits are available, we have marked it with a smaller black dot. To make the transition boundary at $\lambda = \frac{1}{4\pi}$ clear, we have used transseries parameter $\tsgm2 = 2\pi\rmi\, \rme^{2\pi\rmi N}$ \textit{everywhere} on the $(N,\hbar)$-plane of \figref{fig:agreement_N_hbar}. In particular, note how the results cease to match once we cross the diagonal line (this will soon be identified with the Stokes line). Due to numerical instabilities in the calculation, our results are more solid for bigger rather than smaller $\hbar$. 

\begin{table}[t!]
\centering
\ra{1.3}
\begin{tabular}{lllll}
\toprule
\multicolumn{1}{c}{$N$} & \multicolumn{1}{c}{$\hbar$} & \multicolumn{1}{c}{$\hFPtwo$} & \multicolumn{1}{c}{$\CS_0 \hFn{0}$} & \multicolumn{1}{c}{$\CS_0 \hFn{0} + \CS_0 \Phi^{(1)}$} \\
\midrule
$\frac{1}{2}$ & $4\pi$ & $-\dblunderline{1.505\,134}{\,688}\,359\,908\,31$ & $-\underline{1.505\,134}\,801$ & $-\uwave{1.505\,134\,688}$ \\
$\frac{3}{4}$ & $6\pi$ & $-\dblunderline{3.346\,528\,67}{2\,942\,8}43\,55$ & $-\underline{3.346\,528\,67}3\,077\,0$ & $-\uwave{3.346\,528\,672\,942\,8}$ \\
$\frac{2}{3}$ & $8\pi$ & $-\dblunderline{3.622\,022\,579\,02}{3\,56}9\,86$ & $-\underline{3.622\,022\,579\,02}1\,505$ & $-\uwave{3.622\,022\,579\,023\,56}$ \\
$\frac{5}{4}$ & $9\pi$ & $-\dblunderline{8.494\,305\,311\,610}{\,883\,1}2$ & $-\underline{8.494\,305\,311\,610}\,900\,2$ & $-\uwave{8.494\,305\,311\,610\,883\,1}$ \\
\bottomrule  
\end{tabular}
\caption{Comparison of the exact $\hFPtwo$, the perturbative resummation, and the perturbative plus one-instanton resummation, for several values of $N$ and $\hbar$ with $\lambda < \frac{1}{4\pi}$. The value of $\tsgm2$ is that in \eqref{eq:conjectured_tilde_sigma2}. Common digits are marked with straight (perturbative) and wavy (perturbative plus one-instanton) underlines (all displayed digits are stable).}
\label{tab:exact_vs_perturbative_vs_oneinstanton_left}
\end{table}

\begin{figure}[t!]
\begin{center}
\includegraphics[width=\textwidth]{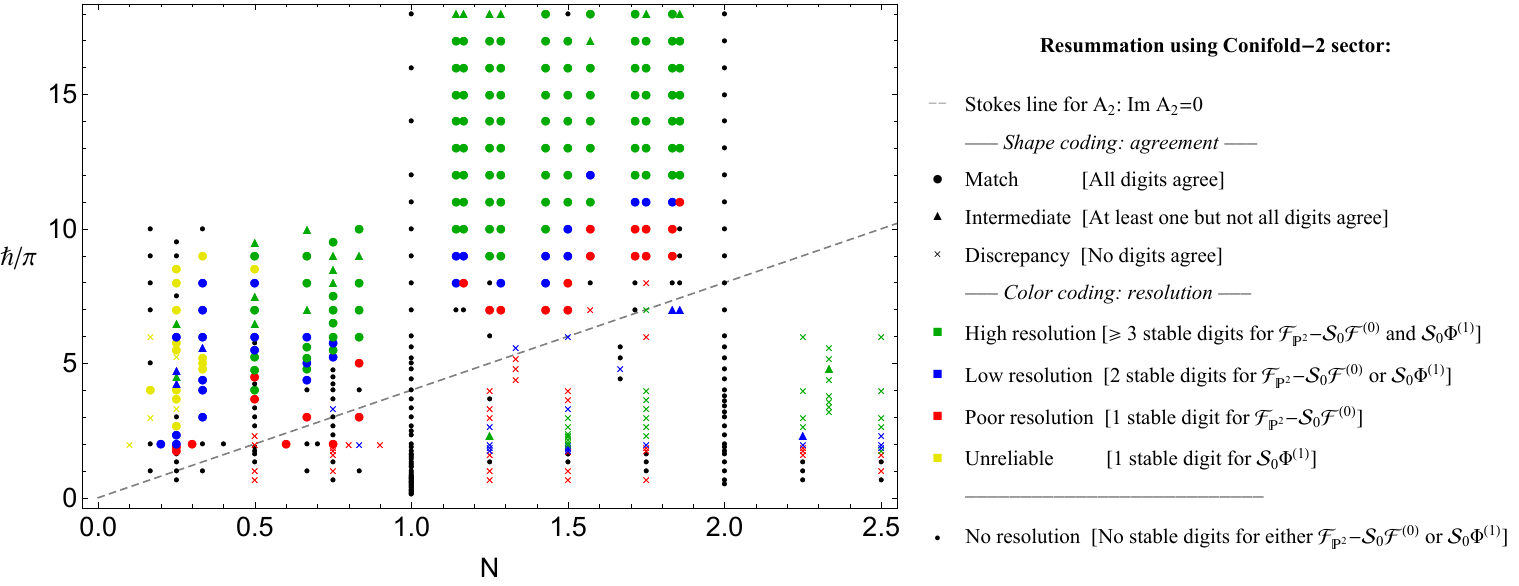}
\end{center}
\caption{This plot shows the agreement between $\hFPtwo - \CS_0 \hFn{0}$ and the one-instanton contribution in the right-hand side of equation \eqref{eq:diff_match_1instA2_v2}. We use $\tsgm2 = 2\pi\rmi \left( \rme^{2\pi\rmi N}-1 \right)$ everywhere, as in \eqref{eq:conjectured_tilde_sigma2} for $\lambda < \frac{1}{4\pi}$, to contrast how this is only correct to the left of the Stokes line. The coding is as in \figref{fig:agreement_N_hbar}. Complete agreement is indicated with a circle. The triangle indicates disagreement in some of the digits and the cross means total disagreement. Green means that at least 3 stable digits are available for comparison; blue indicates 2 stable digits for either quantity, and red means only 1 stable digit for $\hFPtwo - \CS_0 \hFn{0}$. Yellow is the least reliable, with only 1 stable digit for the one-instanton contribution. The dashed line is the Stokes line $\im A_2 = 0$.}
\label{fig:agreement_N_hbar_left_of_Stokes_line}
\end{figure}

Complementarily, we can look at points where $\lambda < \frac{1}{4\pi}$. As before, we show the agreement between $\hFPtwo$ and $\CS_0 \hFn{0} + \CS_0 \Phi^{(1)}$ in Table~\ref{tab:exact_vs_perturbative_vs_oneinstanton_left}. The full $(N,\hbar)$-plane is then displayed in \figref{fig:agreement_N_hbar_left_of_Stokes_line}, where we have used $\tsgm2 = 2\pi\rmi \left(\rme^{2\pi\rmi N}-1\right)$ everywhere so that the line $\lambda = \frac{1}{4\pi}$ stands out. 

As advertised, the origin of the borderline at $\lambda = \frac{1}{4\pi}$ is the Stokes line defined by the relevant instanton action $A_2$, that is,
\begin{equation}
\im A_2 = 0 \quad \text{at} \quad \lambda = \frac{1}{4\pi}. 
\end{equation}
\noindent
To the left and to the right of this line, the value of the transseries parameter changes because the asymptotics of our function $\hFPtwo$ changes. This is the familiar Stokes phenomenon. For resurgent systems, Stokes phenomenon may be described in full detail using the Stokes automorphism (see, \textit{e.g.}, \cite{abs16}). In simple terms, this is a map relating the transseries resummations on both sides of the Stokes line. Since resummations can only differ by the value of the transseries parameter, $\sigma$, the Stokes automorphism dictates how $\sigma$ changes across the Stokes line. In the simplest possible situations, one has the following relation
\begin{equation}
\label{eq:stokes_transition}
\CS_{0^-} F(\sigma) = \CS_{0^+} F(\sigma + S_1),
\end{equation}
\noindent
where $S_1$ is precisely the Stokes constant for the problem. Considering our specific case, equation \eqref{eq:conjectured_tilde_sigma2} tells us that 
\begin{equation}
\tsgm2 \left( \lambda > \frac{1}{4\pi} \right) = \tsgm2 \left( \lambda < \frac{1}{4\pi} \right) + 2\pi\rmi.
\end{equation}
\noindent
Using the original definition $\tsgm2 = \frac{2\pi\rmi}{\stk{A_2}}\, \sgm{A_2}$, this immediately translates to
\begin{equation}
\label{eq:transseries_parameter_stokes_jump}
\sgm{A_2} \left( \lambda > \frac{1}{4\pi} \right) = \sgm{A_2} \left( \lambda < \frac{1}{4\pi} \right) + \stk{A_2},
\end{equation}
\noindent
which is exactly the Stokes phenomenon that equation \eqref{eq:stokes_transition} describes. In conclusion, this implies the transseries resummation at leading order is completely understood. 

\begin{table}[t!]
\centering
\ra{1.3}
\begin{tabular}{cccll}
\toprule
\multicolumn{1}{c}{$N$} &\multicolumn{1}{c}{$\hbar$} & \multicolumn{1}{c}{Pad\'e order} & \multicolumn{1}{c}{$\hFPtwo- \CS_0 \hFn{0}$} & \multicolumn{1}{c}{$|\CS_0 \Phi^{(1)}(\tsgm2=1)|$} \\
\midrule
\multirow{3}{*}{$1$} & \multirow{3}{*}{$10\pi$} & $100$ & $+3.254\cdot 10^{-21}$ & \multirow{3}{*}{$3.924\cdot 10^{-15}$} \\
& & $104$ & $+1.438 \cdot 10^{-20}$ & \\
& & $108$ & $-1.880 \cdot 10^{-20}$ & \\
\midrule
\multirow{3}{*}{$2$} & \multirow{3}{*}{$17\pi$} & $100$ & $+4.901\cdot 10^{-30}$ & \multirow{3}{*}{$4.068\cdot 10^{-24}$} \\
& & $104$ & $+3.113 \cdot 10^{-29}$ & \\
& & $108$ & $-5.031 \cdot 10^{-30}$ & \\
\bottomrule  
\end{tabular}
\caption{Values of the difference $\hFPtwo - \CS_0 \hFn{0}$, above the Stokes line and at integer values of $N$. The numerical instability of the resummations, at various (diagonal) Pad\'e orders, and their being much smaller than $|\CS_0 \Phi^{(1)}(\tsgm2=1)|$, are compatible with the prediction that the perturbative resummation is \textit{exact} for those points.}
\label{tab:perturbative_is_exact}
\end{table}

Finally, there is an interesting consequence to the functional form of $\tsgm2$ when $\lambda < \frac{1}{4\pi}$: if $N\in \BN$ the transseries parameter \textit{vanishes}. This implies that, in these cases, the perturbative resummation is \textit{exact},
\begin{equation}
\hFPtwo \equiv \CS_0 \hFn{0} \quad \text{for} \quad \lambda < \frac{1}{4\pi} \quad \text{and} \quad N\in\BN.
\end{equation}
\noindent
As one tries to check this statement back in \figref{fig:agreement_N_hbar_left_of_Stokes_line}, we see that the points with $N=1$ and $N=2$ above the Stokes line are marked with black dots. This means that $\hFPtwo - \CS_0 \hFn{0}$ has no stable digits. The reason for this is that either the resummation is not accurate enough to display the difference, or that the difference is supposed to be zero.  In order to distinguish between the two cases, and check that indeed $\tsgm2 = 0$, we consider the order of magnitude of $\hFPtwo - \CS_0 \hFn{0}$ versus that of $\CS_0 \Phi^{(1)}$. If the former is much smaller than the latter, we can be confident that $\CS_0 \Phi^{(1)}$ should not appear in the resummation. Table~\ref{tab:perturbative_is_exact} shows a couple of examples where we have displayed the resummations at different Pad\'e orders, illustrating the lack of numerical stability, and also the value $|\CS_0 \Phi^{(1)}|$ with $\tsgm2 =1$, to see that it is indeed much bigger than $\hFPtwo - \CS_0 \hFn{0}$. 

\begin{figure}[t!]
\begin{center}
\includegraphics[width=0.45\textwidth]{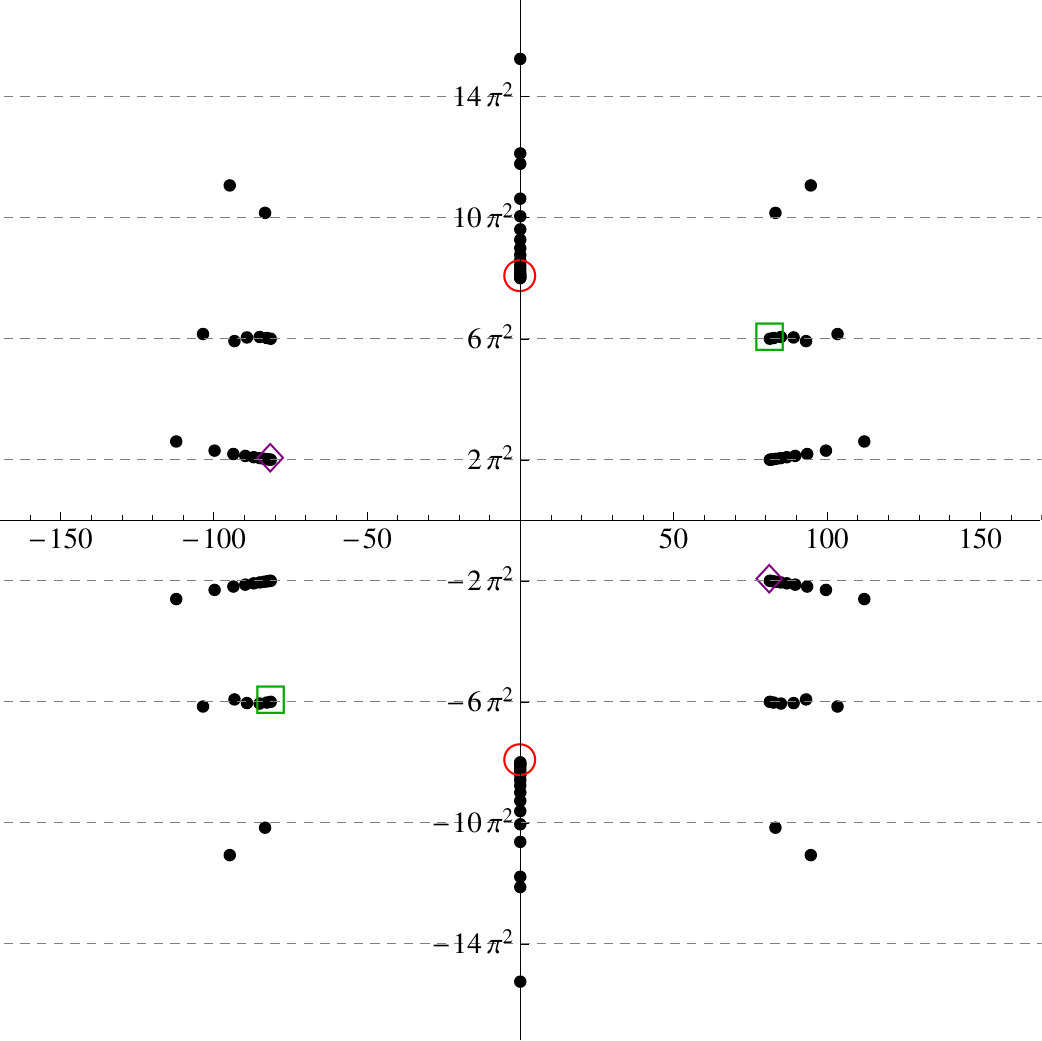}
\hspace{0.05\textwidth}
\includegraphics[width=0.45\textwidth]{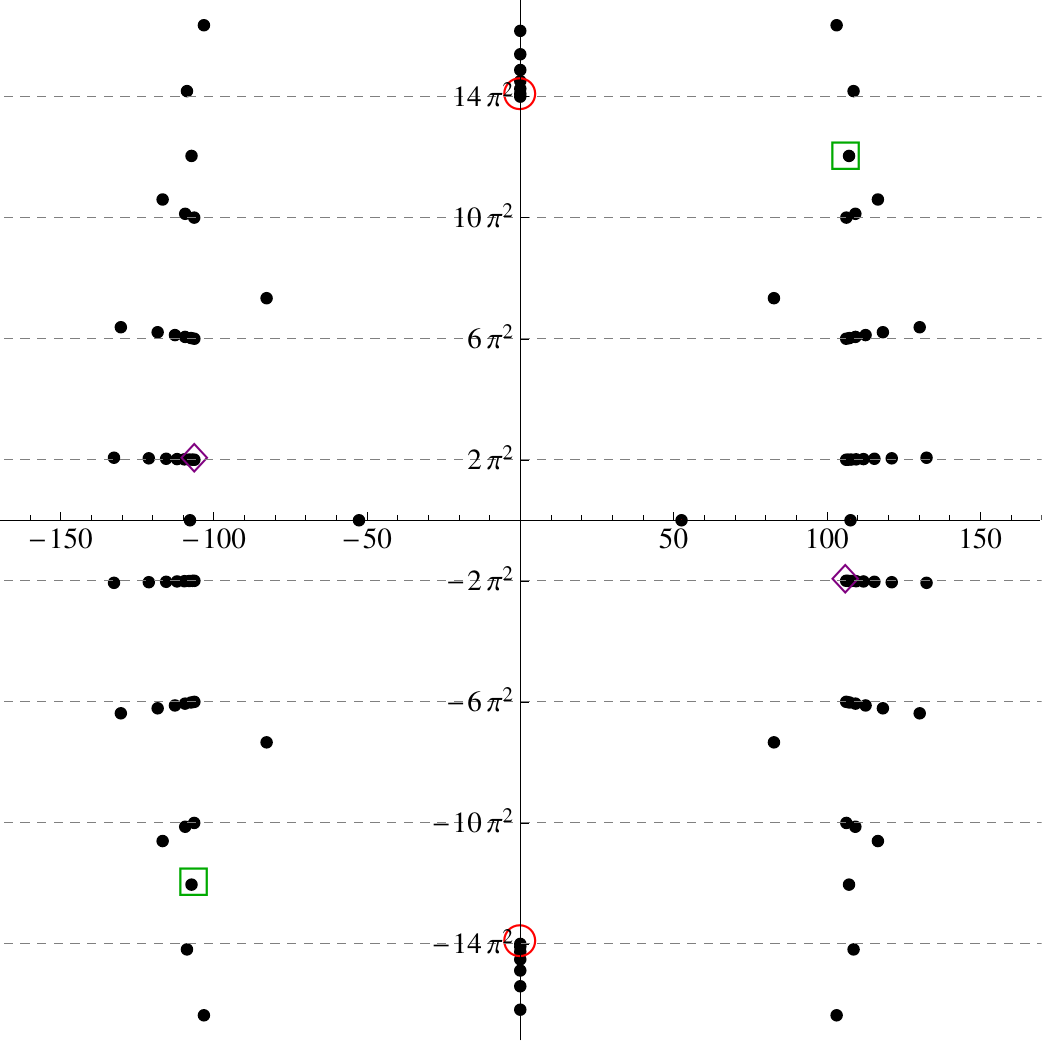}
\end{center}
\caption{Borel singularities yield information on instanton sectors in the transseries. We plot poles of the Pad\'e approximant to the Borel transform of $\hFn{0}$, for $\lambda = \frac{1}{\pi}$ (left) and $\lambda=\frac{7}{4\pi}$ (right). The red circle shows $A_1$, the green square $A_2$, and the purple diamond $A_\txK$. We can see the tower of poles below and above $A_\txK$, and how more and more of them appear as $\lambda$ increases.}
\label{fig:borel_plane_2}
\end{figure}

\begin{figure}[t!]
\begin{center}
\includegraphics[width=0.45\textwidth]{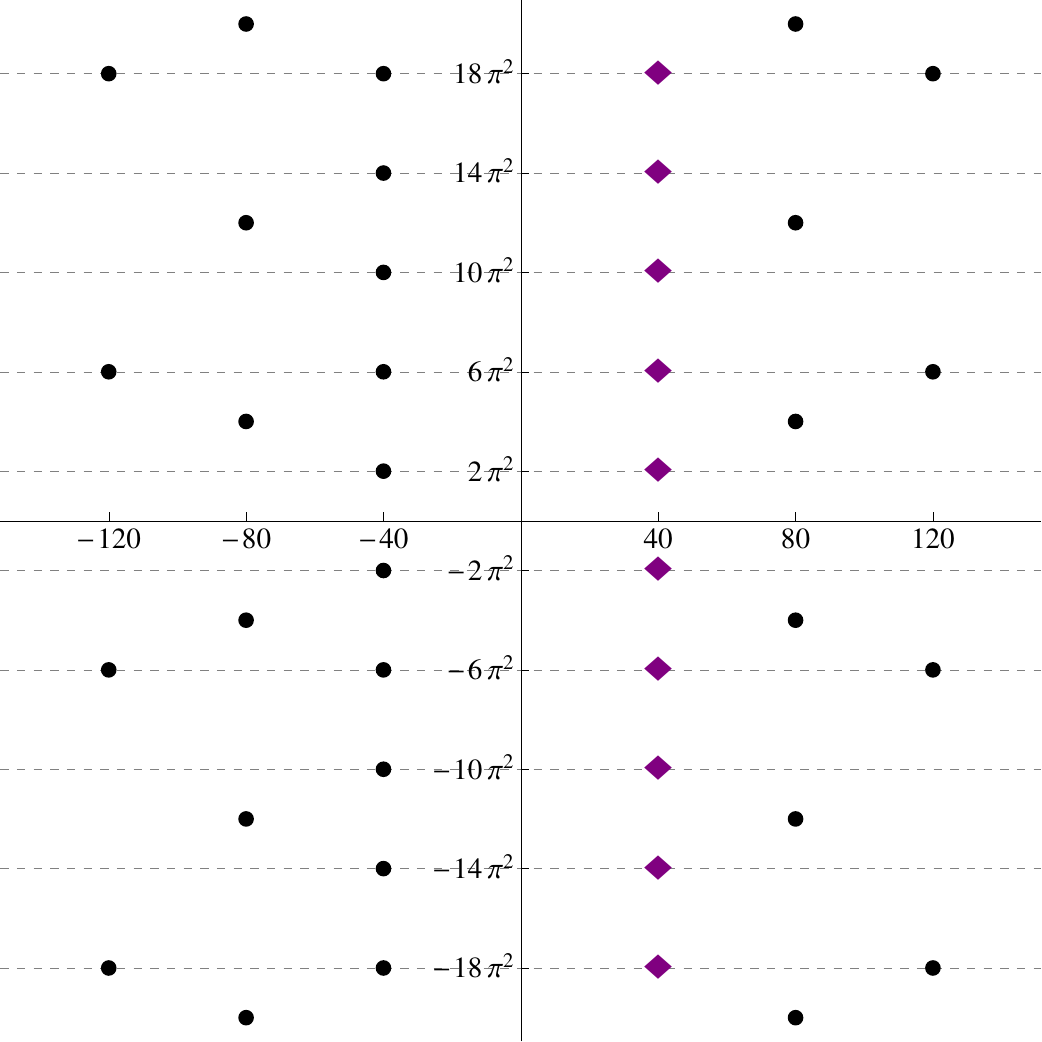}
\end{center}
\caption{The Borel plane for the perturbative free energy of the resolved conifold, for $2\pi t = 40-2\pi^2\rmi$. The singularities (poles) are located at $\pm n \left(2\pi t + 4\pi^2\rmi\, m\right)$, for $m\in\BZ$ and $n\in \BN_{\geq 1}$. The purple diamonds mark the poles with the least non-zero and positive real part, $2\pi t + 4\pi^2\rmi\, m$.}
\label{fig:resolved_conifold_borel_plot}
\end{figure}

In view of the analysis above, our choice of the transseries sector associated to $A_2$ (while setting the other ones to zero), as well as the value of the corresponding transseries parameter \eqref{eq:conjectured_tilde_sigma2}, are justified by the highly non-trivial numerical agreement and by the simplicity of the resulting picture. In particular, the existence of a Stokes line precisely at $\im A_2 = 0$, and the expected Stokes phenomenon when we cross it, \eqref{eq:stokes_transition} and \eqref{eq:transseries_parameter_stokes_jump}, give further evidence that our transseries \textit{ansatz} is correct. However, as described in appendix~\ref{sec:relations_sectors}, we can achieve the same match in the resummation for the sectors of $A_\txK$ and $A_3$, albeit at the cost of much more complicated transseries parameters---as can be seen in \eqref{eq:conjectured_tilde_sigmaK} and \eqref{eq:conjectured_tilde_sigma3}. But then this choice boils down to understanding why the transseries parameter takes the value it does. In particular, it would be interesting to be able to \textit{predict} the value of the Stokes constant, akin to what was done in \cite{m08, csv15}. Unfortunately, we only have a conjectural explanation in the case of $A_\txK$.

When considering the large-radius instanton action, $A_\txK$, other actions, of the form $A_\txK + 4\pi^2\rmi\, m$ with integer $m$, become visible in the Borel plane. We show this in \figref{fig:borel_plane_2}, for two values of $\lambda$. The allowed range of $m$ seems to increase as $\lambda$ grows (towards the large-radius point) since more and more points accumulate there, but because the number of poles is fixed by the finiteness of our current data, we cannot know whether this tower of poles is infinite or not. This picture of a tower of poles is not new: it was seen before in the context of strings on the resolved conifold \cite{ps09}, ABJM gauge theory \cite{gmz14}, and Chern--Simons gauge theory on lens spaces \cite{ars14}. One may wonder how generic will this property be, within the topological string setting.

The resolved conifold is a very well understood geometry and its resurgent properties have been checked in detail \cite{ps09}. A resummation of the perturbative free energy was performed in \cite{ho15} and shown equal to the exact free energy (calculated from Chern--Simons theory) for physical values of $(N,g_{\text{s}})$. There are several instanton actions, arising from either a polylogarithm contribution to the free energy, or from the constant map term. They are, respectively, \cite{ps09}
\begin{align}
A_m &= 2\pi t + 4\pi^2\rmi\, m, \quad m\in \BZ, \\
A_{\text{cm}} &= 4\pi^2 \rmi,
\end{align}
\noindent
where $t = \rmi g_{\text{s}} N$ is the 't~Hooft coupling. We can try to connect this picture to the ones in \figref{fig:borel_plane_2}, by letting $2\pi t \equiv -A_\txK = \re(-A_\txK) - 2\pi^2\rmi$, and forgetting about $A_{\text{cm}}$. In this case the Borel plane for the resolved conifold is shown in \figref{fig:resolved_conifold_borel_plot}. The Borel singularities of the resolved conifold are pole-like rather than logarithmic and, as such, there are no branch cuts emanating from the locations of the instanton actions and the overall picture is cleaner than that for local $\BP^2$. Notice how the infinite column immediately to the right of the imaginary axis, marked as purple diamonds, resembles the poles for local $\BP^2$ in \figref{fig:borel_plane_2}. For real $g_{\text{s}}$ and whenever $F_{\txRC} \neq \CS_0 \Fng{0}{\txRC}$ we should expect that the first tower of poles provides the first corrections to the difference, just like for local $\BP^2$. But in this case, one should sum over all poles, each with its own transseries parameter, $\sgm{m}$. The result would be
\begin{equation}
\text{one-inst.} = \sum_{m\in\BZ} \frac{\sgm{m}}{g_{\text{s}}}\, \rme^{-\frac{A_m}{g_{\text{s}}}} \left( \FnAg{1}{A_m}{\txRC,0} + g_{\text{s}} \FnAg{1}{A_m}{\txRC,1} \right) = \sum_{m\in\BZ} \tsgm{m}\, \rme^{-\frac{2\pi t + 4\pi^2\rmi m}{g_{\text{s}}}} \left( \frac{1}{2\pi^2} + \frac{\rmi t -2\pi m}{\pi g_{\text{s}}} \right),
\end{equation}
\noindent
where we have used the explicit formulae for the $\FnAg{1}{A_m}{\txRC,g}$ from \cite{ps09}, which vanish if $g\geq2$. This expression can be rewritten as a single one-instanton contribution, say for $m=0$, but with an effective transseries parameter, $\tsgm{\txeff}$, given by
\begin{equation}
\text{one-inst.} = \frac{\sgm{\txeff}}{g_{\text{s}}}\, \rme^{-\frac{A_0}{g_{\text{s}}}} \left( \FnAg{1}{A_0}{\txRC,0} + g_{\text{s}} \FnAg{1}{A_0}{\txRC,1} \right) = \tsgm{\txeff}\, \rme^{-\frac{2\pi t}{g_{\text{s}}}} \left( \frac{1}{2\pi^2} + \frac{\rmi t}{\pi g_{\text{s}}} \right),
\end{equation}
\noindent
so that one has
\begin{equation}
\tsgm{\txeff} = \left( \frac{1}{2\pi^2} + \frac{\rmi t}{\pi g_{\text{s}}} \right)^{-1} \sum_{m\in\BZ} \tsgm{m}\, \rme^{-\frac{4\pi^2\rmi m}{g_{\text{s}}}} \left( \frac{1}{2\pi^2} + \frac{\rmi t -2\pi m}{\pi g_{\text{s}}} \right).
\end{equation}
\noindent
It might be possible that this kind of effect would be at work for local $\BP^2$, producing an effective transseries parameter $\tsgm{\txK}$ of the type that we see in equation \eqref{eq:conjectured_tilde_sigmaK} of appendix \ref{sec:relations_sectors}.

\section{Discussion}\label{sec:discussion}

This work shows how the resummation of the resurgent transseries for the free energy of topological strings on the local $\BP^2$ toric CY threefold matches with high precision, after the occurrence of Stokes phenomenon, analytical results obtained via quantization of the mirror curve. This implies in particular that the nonperturbative definition of \cite{ghm14a, mz15} is compatible with the nonperturbative structure obtained in \cite{cesv13, cesv14} from the BCOV holomorphic anomaly equations, \textit{i.e.}, from the underlying closed string field theory of the topological string. It is worth pointing out that our calculations also confirm some general implications of the conjectures put forward in \cite{ghm14a, mz15}. According to these conjectures, the Borel resummation of the string perturbative series and the fermionic spectral traces can only differ in exponentially small quantities in the string coupling constant, which we have explicitly verified. Finally, our results further validate the resurgent transseries techniques of \cite{cesv13, cesv14} as a powerful method to construct nonperturbative observables in topological string theory. Given the range of applicability of the holomorphic anomaly equations, it seems likely that these methods may be extended towards higher-genus mirror curves, beyond the realm of toric threefolds, or even towards the compact case. Some preliminary steps towards higher-genus examples (the ABJM ``slice'' of the local $\BP^1 \times \BP^1$ geometry) were already given in \cite{dmp11}, and towards the (quintic) compact case in \cite{csv16}, albeit both cases promise to be much harder computationally than our present case of local $\BP^2$.

Still, on our current results there are some improvements and loose ends which might still be addressed. One issue deals with the implementation of the numerical resummation. This is a somewhat sensitive procedure which not always leads to stable results, and there is certainly room for improvement on our numerics. From the results in \cite{cesv14} we only have finite data for both perturbative and one-instanton sectors. These data could be improved, albeit having fewer coefficients $\Fng{1}{g}$ is not as crucial as counting on many coefficients $\Fng{0}{g}$. This is because the difference $\hFPtwo - \CS_0 \hFn{0}$ can be quite small, but we need enough stable digits to match against the resummation of the one-instanton sector. The Borel--Pad\'e resummation method is stable most of the time, but not in every case. In this way, our approach to determining stable digits for $\hFPtwo - \CS_0 \hFn{0}$ was based on performing a couple of resummations with different Pad\'e orders, and then only keeping unvarying digits. But we cannot rule-out the existence of other, more efficient, numerical methods. Note that the size of $\hFPtwo - \CS_0 \hFn{0}$ is of order $\rme^{-A/g_{\text{s}}}$. If $g_{\text{s}}$ is not small, this magnitude is big but the resummation $\CS_0 \hFn{0}$ is less tame and the Borel--Pad\'e has to work harder. For $g_{\text{s}}$ small we have the opposite behavior.

Note that improving on available data and numerical methods is not only interesting from the standpoint of having better numerics as compared to the ones in the present paper. It is also quite interesting from the standpoint of enlarging the transseries resummations to, say, negative or even complex values of $N$ (in the spirit of what was done in \cite{csv15} within the matrix model context). Due to Stokes phenomenon, such extensions eventually reach anti-Stokes regions where instantons assume dominance within the transseries and one requires greater amounts of data and sharper numerical precision to carry this through.

\begin{figure}[t!]
\begin{center}
\includegraphics[width=0.45\textwidth]{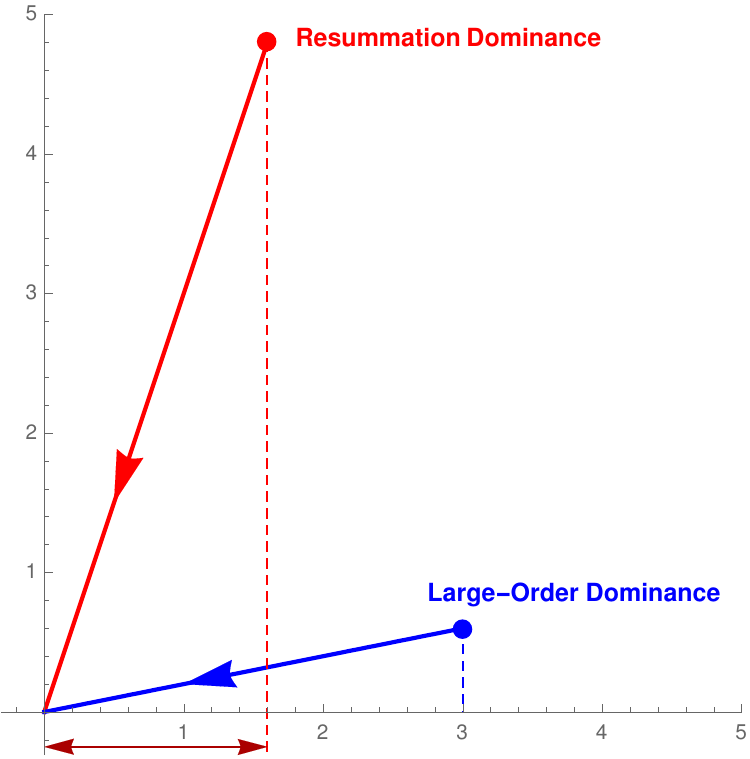}
\end{center}
\caption{Large-order dominance versus resummation dominance, on the Borel plane.}
\label{fig:largeorderVSresummation}
\end{figure}

Still within the realms of our precision, one interesting conclusion may be drawn. As mentioned early-on, our match is also validating the semi-classical data in the topological-string transseries of \cite{cesv13, cesv14, c15}. Let us explain what we mean by this. The acquired knowledge and construction we have for the transseries of the local $\BP^2$ free energy in \cite{cesv14} was obtained via (resurgent) large-order analysis. Iteratively, this type of analysis probes Borel singularities closest to the origin of the Borel plane. But resummation probes something different: it probes projections of these values upon the real axis, and resummation ``dominance'' might very well correspond to \textit{different} singularities as compared to large-order ``dominance''. This is illustrated in \figref{fig:largeorderVSresummation} (see also \cite{bpv78a, bpv78b}). In the present paper the instanton singularities uncovered in \cite{cesv14} via large-order ``dominance'' were shown to be precisely the ones required for resummation, thus validating the nonperturbative (instanton) semi-classical data from \cite{cesv14} as the correct one.
 
In this paper, we have ``decoded'' the nonperturbative information in the fermionic spectral traces of \cite{ghm14a} in terms of the transseries obtained in \cite{cesv13,cesv14}. However, it should also be possible to \textit{derive} this transseries by using explicit expressions for the fermionic spectral traces. For example, in \cite{mz15} a matrix integral representation of $Z_{\mathbb{P}^2}(N, \hbar)$ was given, which was instrumental in order to test the asymptotic expansion \eqref{logz-exp}. In addition, the conjecture of \cite{ghm14a} gives an exact expression for $Z_{\mathbb{P}^2}(N, \hbar)$ as a Laplace transform of the grand potential of the CY, which includes nonperturbative effects. It would be interesting to use these expressions to calculate the transseries from first principles. Such a calculation would lead in particular to a derivation of the transseries parameter \eqref{eq:conjectured_tilde_sigma2}. However, it is important to note that the nonperturbative effects in the Laplace transform formula of \cite{ghm14a} do not have the standard form of a transseries in $g_{\text{s}}$, since they involve trigonometric functions of $1/g_{\text{s}}$. The presence of these functions might be related to the appearance of the towers of Borel poles, as suggested by the calculations in \cite{h15a}.      

A more general issue raised by our analysis is the following: what is the geometric origin of the instanton corrections which appear in the transseries of \cite{cesv13,cesv14}? In standard superstring theory, we expect corrections of this type to come from D-branes or membranes. In our case, since the instanton actions correspond to periods of the CY, we expect them to be due to topological D-branes wrapping cycles of the CY. It would be interesting to test this expectation in detail. 

Finally, one could use the techniques developed in this paper to test whether other proposals in the literature for the nonperturbative completion of topological string theory can be ``decoded'' in terms of resurgent transseries. Unfortunately, many of these proposals are not precise enough to give concrete numbers for the topological-string partition function. In that sense, the proposal of \cite{ghm14a} is quite effective, since the fermionic spectral traces can be computed with relative ease. One other proposal where the ideas of this paper could be applied is the large $N$ duality with Chern--Simons theory \cite{gv98b, akmv02, bb15}, which holds for particular toric backgrounds. In this case, the nonperturbative topological-string partition function is identified with the Chern--Simons partition function on a three-manifold, and one could compare it with the resummation of the relevant transseries, as we have done in this paper. Further, when focusing upon a particular geometry, say local $\BP^1 \times \BP^1$, two very interesting questions immediately emerge. First, one may ask whether the spectral theory and the Chern--Simons nonperturbative completions yield the same result---and if not, in what do they differ. Second, if indeed these definitions do differ nonperturbatively, one may ask whether the very same transseries\footnote{A transseries obtained using the methods of \cite{cesv13} applied to the local $\BP^1 \times \BP^1$ geometry. Note that we already know that local $\BP^1 \times \BP^1$ will have complex instantons with similar flavor to our present example, as shown in \cite{gmz14}.},  with different choices for its parameters, may account for both results---and if so, how do the resulting semi-classical ``decodings'' differ. These are fascinating questions we hope to return to in the near future.

\acknowledgments
We would like to thank Alba Grassi, Yasuyuki Hatsuda and Szabolcs Zakany for discussions. RS would further like to thank the University of Geneva for extended hospitality, where a part of this work was conducted. The research of RCS and RS was partially supported by the FCT-Portugal grants EXCL/MAT-GEO/0222/2012, UID/MAT/04459/2013 and PTDC/MAT-GEO/3319/2014. The research of MM was supported in part by the Fonds National Suisse, subsidies 200021-156995 and 200020-141329. The research of MM and RS was partially supported by the Swiss-NSF grant NCCR 51NF40-141869 ``The Mathematics of Physics'' (SwissMAP).

\newpage

\appendix

\section{The One-Instanton Sector at Large-Radius}\label{sec:checks_1instanton_AK}

For the example of local $\BP^2$, it was shown in \cite{cesv14} that there is an action associated to the K\"ahler modulus, given by $A_\txK = -2 \pi t $, which dominates the large-order growth of perturbation theory near the large-radius point ($z=0$). However, its corresponding sector in the transseries, \textit{i.e.}, the associated coefficients, were not addressed in detail in \cite{cesv14} and we shall do so now. 
In particular, in this appendix we will show that the holomorphic limit of the free energies (in the large-radius frame) satisfies a particular instance of equation \eqref{eq:generic_holo_limit}, namely
\begin{equation}
\label{eq:holo_limit_AK}
\frac{\stk{A_\txK}}{\rmi\pi}\, \hFnAgA{1}{A_\txK}{0}{A_\txK} = 3\, \frac{A_\txK}{2\pi^2}, \qquad \frac{\stk{A_\txK}}{\rmi\pi}\, \hFnAgA{1}{A_\txK}{1}{A_\txK} = 3\, \frac{1}{2\pi^2}, \qquad \frac{\stk{A_\txK}}{\rmi\pi} \hFnAgA{1}{A_\txK}{g\geq 2}{A_\txK} = 0.
\end{equation}
\noindent
These equations are analogous to the ones for the conifold sectors, except for the factor of $3$ which might be identified with the GV invariant $n^{(1)}_0 = 3$. However, unlike for the conifold sector, we have no analytic derivation for equation \eqref{eq:holo_limit_AK} and we will need to resort to numerical checks. These checks happen to be less well-behaved than for other sectors, as we have obtained slower and more oscillatory convergence.

In practice we take the resurgence formula (see, \textit{e.g.}, \cite{abs16})
\begin{equation}
\Fng{0}{g} \simeq \sum_{h=0}^{+\infty} \frac{\Gamma(2g-1-h)}{A_\txK^{2g-1-h}}\, \frac{\stk{A_\txK}}{\rmi\pi}\, \FnAg{1}{A_\txK}{h}, 
\end{equation}
\noindent
and we evaluate it for different values of $z$, near the large-radius point, and values of $\Szz = \hSzzA{A_\txK}$ as in equation \eqref{eq:Szz_hol_AK}. Considering the large-$g$ limit of the appropriate functions, just as in equation \eqref{eq:glimit_AK_loop_1}, we can check each of the expressions in \eqref{eq:holo_limit_AK}:
\begin{align}
\label{eq:def_Q1_h_g}
\CQ^{(1)}_{h,g} &:= \frac{A_\txK^{2g-1-h}}{\Gamma(2g-1-h)} \left( \hFngA{0}{g}{A_\txK} - \sum_{k=0}^{h-1} \frac{\Gamma(2g-1-k)}{A_\txK^{2g-1-k}}\, \frac{\stk{A_\txK}}{\rmi\pi}\,  \hFnAgA{1}{A_\txK}{k}{A_\txK} \right), \\
\lim_{g\to+\infty}& \CQ^{(1)}_{0,g} = \frac{\stk{A_\txK}}{\rmi\pi}\, \hFnAgA{1}{A_\txK}{0}{A_\txK} = 3\, \frac{A_\txK}{2\pi^2}, \\
\lim_{g\to+\infty}& \CQ^{(1)}_{1,g} = \frac{\stk{A_\txK}}{\rmi\pi}\, \hFnAgA{1}{A_\txK}{1}{A_\txK} = 3\, \frac{1}{2\pi^2}, \\
\lim_{g\to+\infty}& \CQ^{(1)}_{h\geq 2,g} = \frac{\stk{A_\txK}}{\rmi\pi}\, \hFnAgA{1}{A_\txK}{h\geq 2}{A_\txK} = 0.
\label{eq:lim_Q1_hgeq2_g}
\end{align}

\begin{figure}[t!]
\begin{center}
\includegraphics[width=0.46\textwidth]{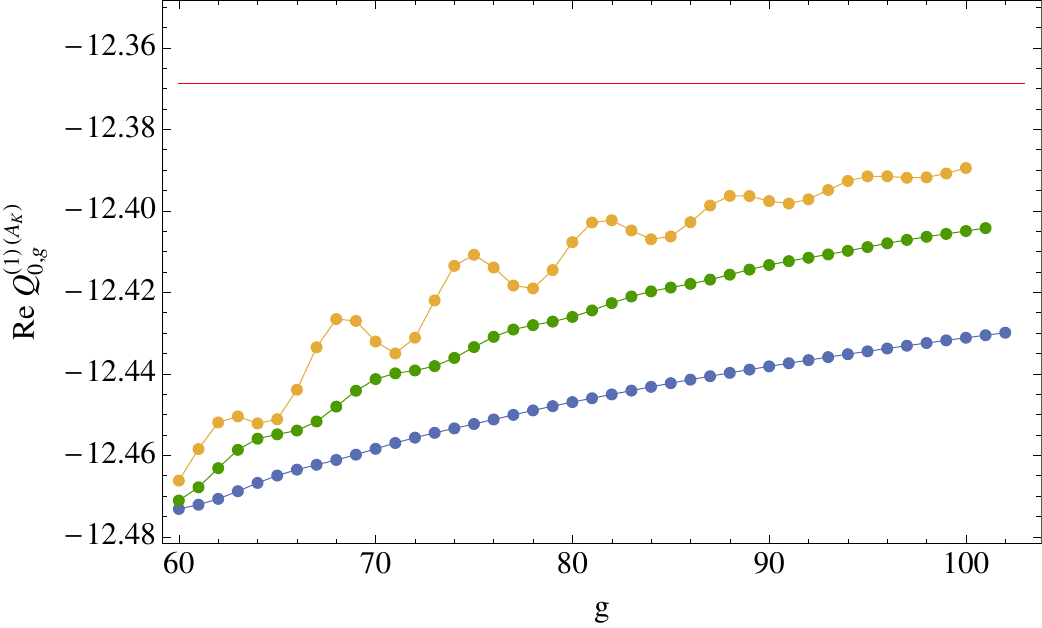}
\hspace{0.05\textwidth}
\includegraphics[width=0.46\textwidth]{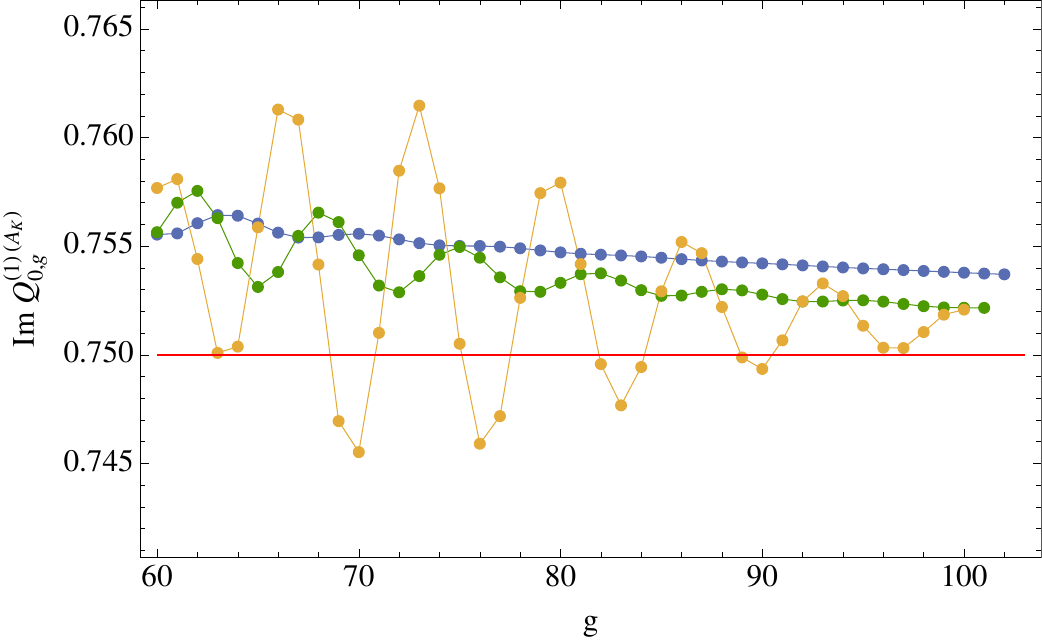}
\includegraphics[width=0.46\textwidth]{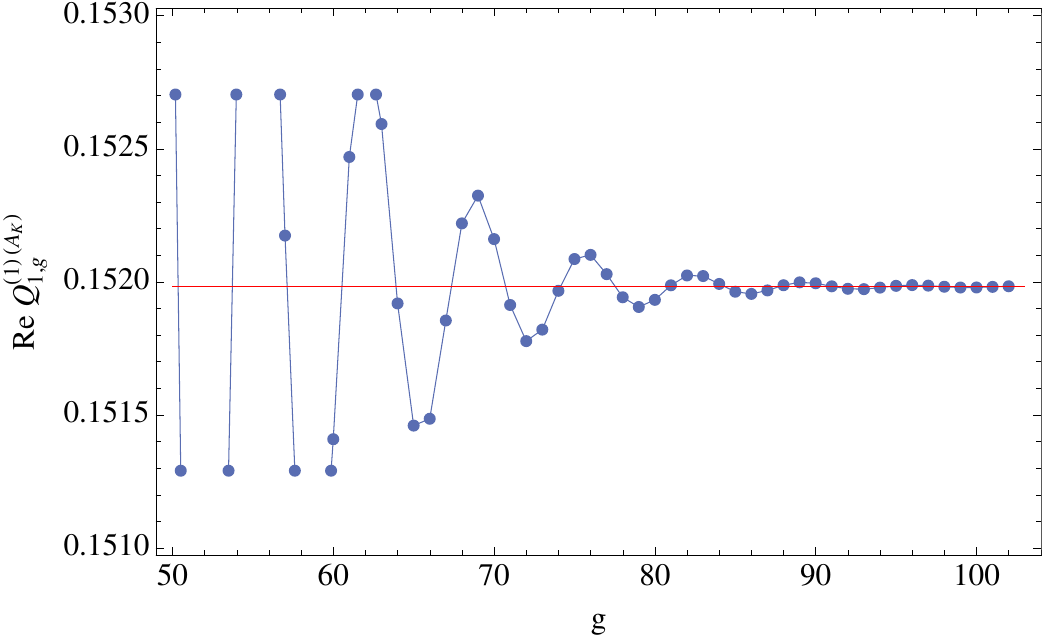}
\hspace{0.05\textwidth}
\includegraphics[width=0.46\textwidth]{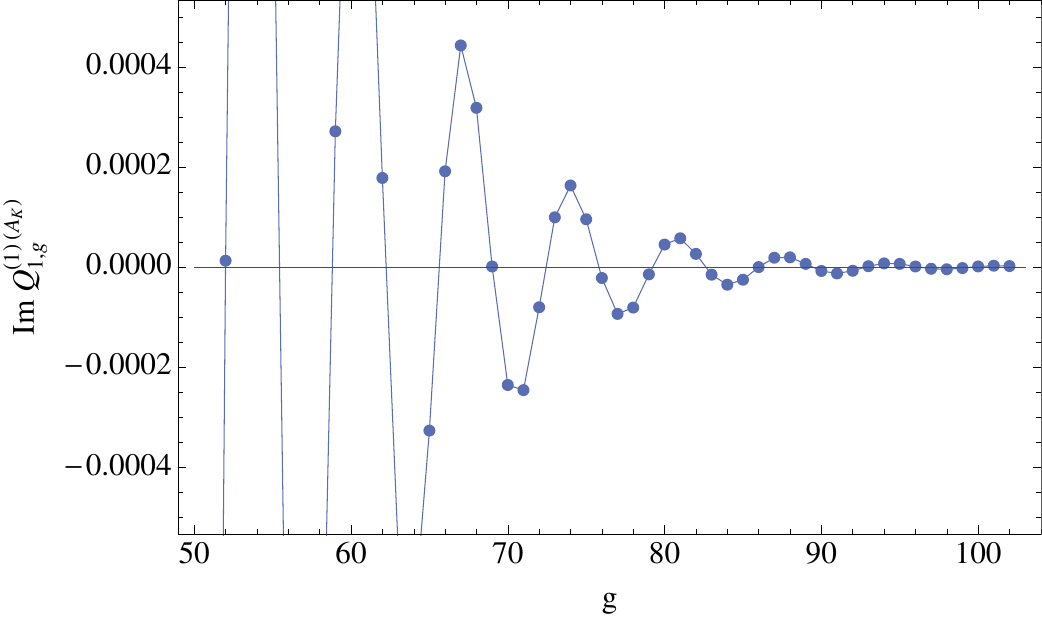}
\includegraphics[width=0.46\textwidth]{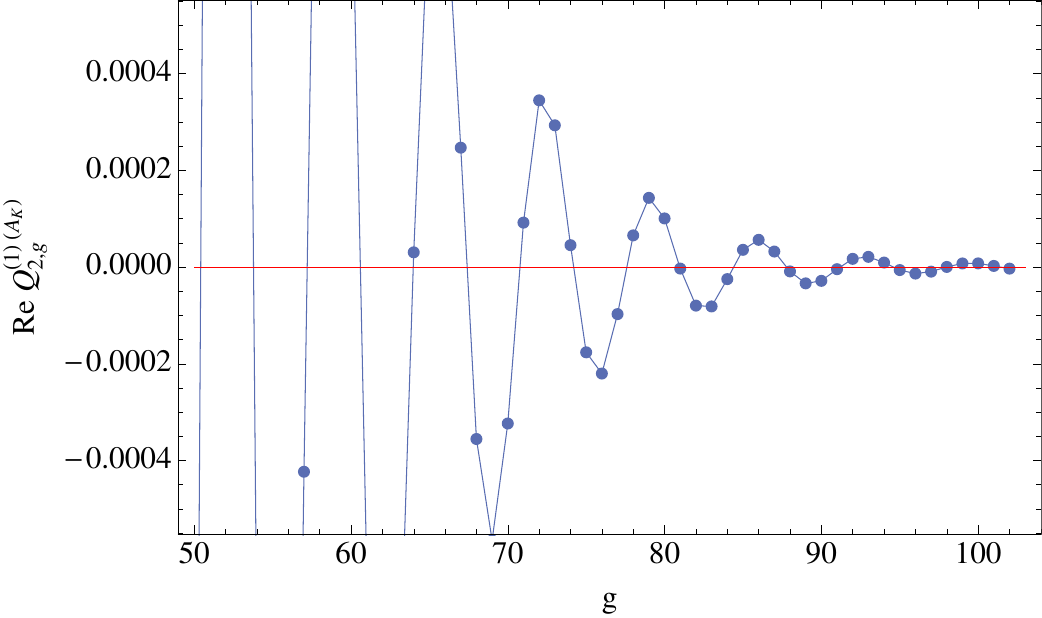}
\hspace{0.05\textwidth}
\includegraphics[width=0.46\textwidth]{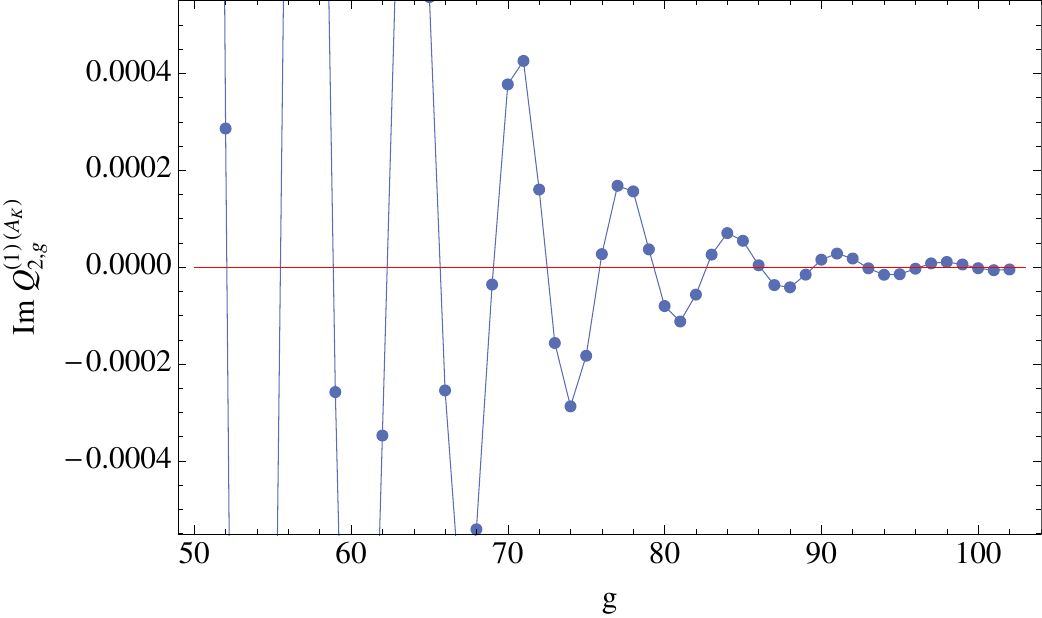}
\end{center}
\caption{Numerical check of equations \eqref{eq:def_Q1_h_g}-\eqref{eq:lim_Q1_hgeq2_g} for a value of the complex modulus $z=-75^{-3}\rme^{-3\pi\rmi/4}$. We show both the real (right) and imaginary (left) parts of the limits. For the first loop ($h=0$, top plots)  we use two Richardson transforms to accelerate convergence, although we can see that the oscillations are large. The other plots do not need convergence acceleration.}
\label{fig:checks_1instanton_AK}
\end{figure}

In the plots of \figref{fig:checks_1instanton_AK} we have fixed the value of $z$ and performed these limits numerically. Accelerating convergence with Richardson extrapolation we can get quite close to the expected results (see, \textit{e.g.}, \cite{abs16} for an introduction to these methods). The numerical results and the corresponding predictions are:
\begin{alignat}{3}
&h=0: \qquad \lim_{g\to+\infty} \CQ^{(1)}_{0,g} = -12.38950 + 0.752078\rmi, \qquad &\frac{3 A_\txK}{2\pi^2} &= -12.36870 + 0.749990 \rmi, \\
&h=1: \qquad \lim_{g\to+\infty} \CQ^{(1)}_{1,g} = 0.1519832 + 1.9 \cdot 10^{-6}\rmi, \qquad &\frac{3}{2\pi^2} &= 0.1519817, \\
&h=2: \qquad \lim_{g\to+\infty} \CQ^{(1)}_{2,g} = -3.2 \cdot 10^{-6} -4.9 \cdot 10^{-6}\rmi, \qquad &\text{pred.} &= 0.
\end{alignat}

\section{Relations between Instanton Transseries Sectors}\label{sec:relations_sectors}

In section~\ref{sec:resummation} we matched the exact topological-string free energy of local $\BP^2$, $\hFPtwo$, with the resummation of the corresponding transseries whose only non-zero transseries parameter (\textit{i.e.}, transseries sector) was that of $A_2$, and according to \eqref{eq:conjectured_tilde_sigma2}. In particular, the value of this parameter, $\sgm{A_2}$, changed in the expected way at the Stokes line, $\im A_2 = 0$ or $\lambda = \frac{1}{4\pi}$. This appendix discusses how one can still find a match between exact values and transseries resummation, if instead we allow only the $A_3$ or the $A_\txK$ sectors to be non-zero. However, in this process the values of the corresponding transseries parameters will not be as simple as $\tsgm{A_2}$ in equation \eqref{eq:conjectured_tilde_sigma2}.

The reason why these other sectors may be used, is that there are \textit{specific relations} between the several instanton actions \cite{cesv14}; namely
\begin{align}
A_2 + A_\txK - A_1 &= 0, \\
A_1 + A_2 + A_3 + 4\pi^2 \rmi &= 0.
\end{align}
\noindent
For real values of $\lambda$, we also have the following identities
\begin{gather}
\re(A_1) = 0, \qquad \re(A_2) = \re(-A_\txK) = \re(-A_3), \\
\im(A_\txK) = 2\pi^2, \quad \im(A_1) = 8\pi^3 \lambda, \quad \im(A_2) = 8\pi^3 \left( \lambda - \frac{1}{4\pi} \right), \quad \im(A_3) = -16\pi^3 \left( \lambda + \frac{1}{8\pi} \right).
\end{gather}
\noindent
These relations then have consequences for the respective free energies, at the one-instanton level, and in the holomorphic limit of the conifold-1 frame. One finds:
\begin{align}
\re\,\thFnAg{1}{A_2}{g} &= -\frac{1}{3}\, \re\,\thFnAg{1}{-A_\txK}{g}, \\ 
\im\,\thFnAg{1}{A_2}{g} &= \frac{1}{\pi}\, \im\,\thFnAg{1}{-A_\txK}{g-1} + \frac{4\pi}{3} \left( \lambda - \frac{1}{4\pi} \right) \im\,\thFnAg{1}{-A_\txK}{g},
\label{eq:imF1A2_imF1AK} \\
\re\,\thFnAg{1}{A_2}{g} &= -\re\,\thFnAg{1}{-A_3}{g}, \\ 
\frac{6}{\pi}\, \im\,\thFnAg{1}{A_2}{g-1} + \left( 1 + 8\pi\lambda \right) \im\, \thFnAg{1}{A_2}{g} &= -\frac{3}{\pi}\, \im\,\thFnAg{1}{-A_3}{g-1} + \left( 1 - 4\pi\lambda \right) \im\,\thFnAg{1}{-A_3}{g},
\label{eq:imF1A2_imF1A3}
\end{align}
\noindent
where we have used the usual notation $\thFnAg{1}{A}{g} := \frac{\stk{A}}{2\pi\rmi} \hFnAgA{1}{A}{g}{A_1}$.

Given all these relations, alongside the matching condition \eqref{eq:diff_match_1instA2_v2} for the $A_2$-sector, we can determine the values of $\tsgm{\txK}$ and $\tsgm{3}$ which make the following analogous conditions hold
\begin{align}
\hFPtwo - \CS_0 \hFn{0} &= \re \left( \tsgm{\txK}\, g_{\text{s}}^{-1}\, \rme^{A_\txK/g_{\text{s}}}\, \CS_0 \thFnA{1}{-A_\txK} \right), \\
\hFPtwo - \CS_0 \hFn{0} &= \re \left( \tsgm{3}\, g_{\text{s}}^{-1}\, \rme^{A_3/g_{\text{s}}}\, \CS_0 \thFnA{1}{-A_3} \right).
\end{align}
\noindent
After defining $\nu = 2\pi \left( \frac{3}{\hbar} + \lambda \right)$, we find
\begin{align}
\tsgm{\txK} &= \begin{cases}
\frac{2\pi\rmi}{3} \left( -1 + \left( 1 + \rme^{-\rmi\hbar} \right) \nu \right), & \lambda > \frac{1}{4\pi}, \\
\frac{2\pi\rmi}{3} \left( \rme^{-2\pi\rmi N} - 1 + \left( \rme^{2\pi\rmi N} - 1 \right) \left( \rme^{-2\pi\rmi N} - \rme^{-\rmi\hbar} \right) \nu \right), & \lambda < \frac{1}{4\pi},
\end{cases}
\label{eq:conjectured_tilde_sigmaK} \\
\tsgm{3} &= \begin{cases}
2\pi\rmi\, \rme^{4\pi\rmi N}\, \frac{ 1 + \left( 1 - 3 \rme^{\rmi\hbar} \right) \nu}{1 + 4 \nu}, & \lambda > \frac{1}{4\pi}, \\
2\pi\rmi\, \rme^{2\pi\rmi N} \left( 1 - \rme^{2\pi\rmi N} \right) \frac{3 \rme^{\rmi\hbar} \nu + \rme^{2\pi\rmi N} \left( 1 + \nu \right)}{1 + 4 \nu}, & \lambda < \frac{1}{4\pi}.
\end{cases}
\label{eq:conjectured_tilde_sigma3}
\end{align}
\noindent
The presence of $\hbar$ arises because equations \eqref{eq:imF1A2_imF1AK} and \eqref{eq:imF1A2_imF1A3} depend on loop-indices $g$ and $g-1$. Furthermore, these relations are not nearly as clean as \eqref{eq:conjectured_tilde_sigma2}, which validates the choice done in the main text of using $A_2$ as the natural sector in the transseries.

\newpage

\bibliographystyle{plain}

\end{document}